\documentclass [12pt]{article}
\usepackage{amsmath,amssymb,cite}

\usepackage{tabularx}
\usepackage[english]{babel}
\usepackage[dvips]{graphicx}
\usepackage{indentfirst}
\usepackage{epsfig}
\numberwithin{equation}{section}

\newcommand{\Trl}{\;\mbox{Tr}_L\;}

\newcommand{\yy}{\hat{\mathcal{Y}}}
\newcommand{\yyd}{\hat{\mathcal{Y}}^{+}}

\begin{document}
\title{Towards Noncommutative Fuzzy QED}
\author{Rodrigo Delgadillo-Blando$^{a,b}$\footnote{Email: rodrigo@synge.stp.dias.ie.}, Badis Ydri$^{c}$\footnote{Email: ydri@physik.hu-berlin.de.}~\footnote{ The work of B.Y is supported by a Marie Curie Fellowship from The Commision of the European Communities ( The Research Directorate-General ) under contract number MIF1-CT-2006-021797. The Humboldt-Universitat Zu Berlin preprint number is HU-EP-06/40.}\\
\\
$^{a}$School of Theoretical Physics, Dublin Institute for Advanced Studies \\Dublin, Ireland.\\
$^{b}$Departamento de F\'{\i}sica, Centro de Investigaci\'on y Estudios Avanzados del IPN\\ M\'exico DF. M\'exico.\\ 
$^{c}$Humboldt-Universitat Zu Berlin, Institut fur Physik\\
Newtonstr.15, 
D-12489 Berlin, Germany}

\maketitle

\begin{abstract}
We study in one-loop perturbation theory noncommutative fuzzy quenched QED$_4$. We write down the effective action on fuzzy ${\bf S}^2\times {\bf S}^2$ and show the existence of a gauge-invariant UV-IR mixing in the model in the large $N$ planar limit. We also give a derivation of the beta function and comment on the limit of large mass of the normal scalar fields. We also discuss topology change in this $4$  fuzzy dimensions arising from the interaction of fields ( matrices ) with spacetime through its noncommutativity.

\end{abstract}

The principal motivation behind noncommutative fuzzy physics \cite{badisth,denjoe,Bal,harald} is the construction of a new nonperturbative method for gauge theories ( commutative and noncommutative ) based on  the fuzzy sphere ${\bf S}^2_N$ and its cartesian products. The actions we obtain on  ${\bf S}^2_N$  are essentially finite dimensional matrix models. The noncommutative Moyal-Weyl spaces are also matrix models not continuum manifolds. They only act on infinite dimensional Hilbert spaces and thus we can use the fuzzy sphere and its cartesian products as  finite dimensional regularizations of these spaces.  
The limit $N{\longrightarrow}\infty $ is the limit of the continuum sphere. The  double scaling noncommutative planar limit of large $R$ ( radius of the sphere ) 
and large $N$ 
keeping ${R^2}/{N}$ fixed  equal to $ {\theta}^2$ is the limit of the noncommutative plane. 

In this article we will illustrate this approach by reviewing the example of noncommutative fuzzy quenched QED$_2$ in which the fuzzy sphere \cite{madore} is the underlying regulator. Then we will generalize the results   to the $4-$dimensional case where the underlying space is  fuzzy ${\bf S}^2\times {\bf S}^2$ \cite{S2S2}. Perturbation theory on fuzzy ${\bf S}^2\times {\bf S}^2$ can be found in the first reference of \cite{uvir}. Quantum fuzzy fermions will be discussed elsewhere \cite{in-progress}. The theories we get by including fermions are the noncommutative fuzzy Schwinger model and noncommutative fuzzy QED$_4$. For noncommutative Moyal-Weyl QED see \cite{recent1,recent,recent3,recent2}. Fuzzy QED as opposed to Moyal-Weyl QED is fully $SO(4)-$invariant and fully finite.

An alternative way of regularizing gauge theories  on the Moyal-Weyl noncommutative space is based on the matrix model formulation of the twisted Eguchi-Kawai model \cite{kawai}. For example a  non-perturbative study of pure two dimensional noncommutative gauge theory was performed in \cite{jun-wolf}. 

However the  advantage of the fuzzy regulator compared to the Eguchi-Kawai models and/or to ordinary lattice prescriptions is that discretization by quantization which leads to noncommutative fuzzy spaces is remarkably successful in preserving symmetries and topological features \cite{fmd1,fmd2}. Most important of all are topological quantities, chiral fermions and supersymmetries which can be formulated in a rigorous way on fuzzy spaces \cite{badisth,denjoe,Bal,harald}. 

The plan of the paper is as follows.

\tableofcontents

\section{Noncommutative fuzzy quenched QED$_2$}

Let us review noncommutative fuzzy quenched QED$_2$.

Noncommutative $U(n)$ gauge theory in two dimensions on the fuzzy sphere ${\bf S}^2_{L+1}$ can be given in terms of three $N\times N$ matrices $X_a$ ( $N=n(L+1)$ ) through the pure $3-$matrix model action ( with $2$ parameters  $\alpha$ and $m$ ) 
\begin{eqnarray}
S=N\bigg[-\frac{1}{4}Tr[X_a,X_b]^2+\frac{2i{\alpha}}{3}{\epsilon}_{abc}TrX_aX_bX_c\bigg]-Nm^2{\alpha}^2 TrX_a^2+\frac{Nm^2}{2c_2} Tr(X_a^2)^2.\label{main2}
\end{eqnarray}
This action is invariant under $1)$ $U(N)$ unitary transformations and  $2)$ $SU(2)$ rotations. The classical absolute minimum of the model is given by the fuzzy sphere configurations \cite{madore}
\begin{eqnarray}
X_a=\alpha L_a {\otimes}{\bf 1}_n\label{fuzz}
\end{eqnarray}
$L_a$ are the generators of spin $\frac{L}{2}$ IRR of $SU(2)$ which satisfy
$[L_a,L_b]=i{\epsilon}_{abc}L_c~ ,~ c_2=\sum_a L_a^2=\frac{L}{2}
(\frac{L}{2}+1)$. The coordinates on the fuzzy sphere ${\bf S}^2_{L+1}$ are defined by 

\begin{eqnarray}
x_1^2+x_2^2+x_3^2=1~,~
[x_a,x_a]=\frac{i}{\sqrt{c_2}}{\epsilon}_{abc}x_c,~x_a=\frac{L_a}{\sqrt{c_2}}.
\end{eqnarray}
Expanding the action (\ref{main2}) around this solution by writing $X_a=\alpha R D_a$ yields $U(n)$ gauge theory  on the fuzzy sphere which is given by

\begin{eqnarray}
S_{L,R}&=&\frac{R^2}{4g^2N}TrF_{ab}^{2}-\frac{R}{2g^2N}{\epsilon}_{abc}Tr\left[\frac{1}{2}F_{ab}^{}A_c-\frac{i}{6}[A_a,A_b]A_c\right]+\frac{2m^2R^2}{g^2N}Tr{\Phi}^2.\label{action2}
\end{eqnarray}
In above $D_a=\frac{1}{R}L_a+A_a$, $F_{ab}=i[D_a,D_b]+\frac{1}{R}{\epsilon}_{abc}D_c$ and $\Phi$ is the covariant scalar field ${\Phi}=\frac{1}{2R}(x_aA_a+A_ax_a)+\frac{A_a^2}{2\sqrt{c_2}}$ where $R$ is the radius of the sphere and $g^2=1/(N^2R^2{\alpha}^4)$ has now the dimension of $(\rm lenght)^{-2}$.  The limit $m{\longrightarrow}\infty$ means that the normal component of $A_a$ ( i.e $\Phi=A_an_a$ ) is $0$.

The  other limit of interest  is a   double scaling noncommutative planar limit of large $R$
and large $L$ taken together restricting the theory in a covariant way around the north pole   and  
keeping ${R^2}/{\sqrt{c}_2}$ fixed  equal $ {\theta}^2$.
The action (\ref{action2}) ( with $m=0$ \footnote{The terms which are proportional to $m^2$ are not needed in this limit.} ) is seen to tend to the action \cite{badis1}
\begin{eqnarray}
S_{\theta}=\frac{{\theta}^2}{8g^2}Tr{\hat F}_{ij}^2=\frac{{\theta}^2}{8g^2}Tr\bigg(i[{\hat D}_i,{\hat D}_j]+\frac{1}{{\theta}^2}{\epsilon}_{ij}\bigg)^2.\label{action1}
\end{eqnarray}
Here ${\hat D}_i=\frac{1}{{\theta}^2}{\hat x}_i+{\hat A}_i$, ${\hat D}_3=\frac{R}{{\theta}^2}$ where $\hat{D}_a=D_a$, $\hat{A}_a=A_a$ and
\begin{eqnarray}
 [{\hat x}_i,{\hat x}_j]=i{\theta}^2{\epsilon}_{ij}~, ~{\hat x}_3=R~,~\hat{x}_a=Rx_a.
\end{eqnarray}
In two dimensions the action (\ref{action1}) is the infinite dimensional matrix model describing $U(n)$ gauge theory on the noncommutative Moyal-Weyl plane \cite{others1}. In this case the trace $Tr$  is an infinite dimensional trace.

The action (\ref{action2}) with the Chern-Simons-like term  and with $m=0$ is precisely what we obtain in the zero-slope limit of the theory of open strings moving in a curved background with ${\bf S}^3$ metric in the presence of a non-zero NS B-field \cite{ars}. 
The action (\ref{action1}) is obtained on the other hand  when open strings are moving in a flat background \cite{sw}. 

As it turns out the path integrals of $U(n)$ models  on the fuzzy sphere ${\bf S}^2_{L+1}$ given by (\ref{main2}) are in one-to-one correspondence with the path integrals of $U(1)$ models on the fuzzy spheres ${\bf S}^2_N$ with $N=n(L+1)$ and thus it is enough to consider only the $U(1)$ case \cite{in-progress}. These $U(1)$ theories are  given by the matrix models (\ref{main2}) or the noncommutative gauge actions (\ref{action2}) with $N=L+1$. In the remainder of this introduction we will  discuss the quantum $U(1)$ gauge theory on the fuzzy sphere ${\bf S}^2_N$. In perturbation theory the quadratic  effective action for the $U(1)$ theory on ${\bf S}^2_N$ given by (\ref{action2})  with the value $m=0$ is found in the continuum limit $N{\longrightarrow}{\infty}$ to be given ( modulo scalar-type terms ) by \cite{ref}

\begin{eqnarray}
{\Gamma}[A]&=& \frac{1}{4g^2}\int
\frac{d{\Omega}}{4{\pi}}F_{ab}^{}\big(1+4g^2\frac{{\Delta}_3}{{\cal L}^2}\big)F_{ab}^{}-\frac{1}{4g^2}{\epsilon}_{abc}\int
\frac{d{\Omega}}{4{\pi}}F_{ab}^{}\big(1+4g^2
\frac{{\Delta}_3}{{\cal L}^2}\big)A_c+....\label{main1}
\end{eqnarray}
${\cal L}^2$ is the Laplacian on the commutative sphere ${\cal L}^2={\cal L}_a^2$, ${\cal L}_a=-i{\epsilon}_{abc}n_b{\partial}_c$. The operator ${\Delta}_3$ is a  function of the Laplacian ${\cal L}^2$ which is defined by its eigenvalues on the spherical harmonics ${Y}_{pm}$ given by $
{\Delta}_3(p)=\sum_{n=2}^{p}1/n$.
 The $1$  in $1+4g^2{\Delta}_3/{\cal L}^2$ corresponds to the classical action whereas ${\Delta}_3/{\cal L}^2$ is the quantum correction. This provides a non-local renormalization of the inverse coupling constant $1/g^2$. We have thus established the existence of a gauge-invariant UV-IR mixing problem in $U(1)$ gauge theory on the fuzzy sphere for $m=0$. Indeed we can immediately see that in the planar limit the eigenvalues of ${\Delta}_3/{\cal L}^2$ behave as $\log p/p^2$ which show a typical singularity at zero momentum associated with the usual UV-IR mixing phenomena \cite{uvir}. In this planar limit we can also show that this singularity at $p{\longrightarrow}0$ is equivalent to a singularity at $\theta{\longrightarrow}0$ in accordance with \cite{min}.

The same result will hold for generic values of the parameter $m$. However we can show that this UV-IR mixing problem is due to the scalar sector of the model in the following sense.  If we decide to quantize the model (\ref{action2}) and then take the limit $m{\longrightarrow}\infty$ and then the limit $N{\longrightarrow}{\infty}$  then one finds that the effective action of the two-dimensional gauge field will be given essentially by the classical action and hence there will be no UV-IR mixing phenomena.  In other words the fuzzy model in this limit is just a fully finite and fully symmetric truncation of the continuum. This complete regularization of the UV-IR mixing through taking a double scaling limit in this particular way happens only in $2$ dimensions with gauge fields \cite{ref}. The origin of the UV-IR mixing in this case seems to lie in the coupling of the $2$ dimensional gauge field to the extra mode present in the model which is the normal scalar component $\Phi$ of $A_a$. This coupling is however unavoidable because the differential calculus on the fuzzy sphere is intrinsically $3-$dimensional. The limit $m{\longrightarrow}\infty$ kills this mode in a covariant way. This perturbative result seems also to be consistent with the $1/N$ expansion of \cite{steinackers2} but not with the full non-perturbative study done using numerical Monte Carlo simulation in \cite{ref1}. So clearly this perturbative picture is not the full story.

A more ( almost non-perturbative ) direct check for the UV-IR mixing in this theory can be given in terms of the effective potential. The quantum minimum is found by considering the configurations 
\begin{eqnarray}
X_a=\alpha {\phi}L_a
\end{eqnarray}
where the order parameter $\alpha \phi$ plays the role of the radius of the sphere. For small values of $m$ the complete one-loop effective potential is given in the large $N$ limit by ( with $\tilde{\alpha}=\sqrt{N}{\alpha}$ ) \cite{ref}
 \begin{eqnarray}
V_{\rm eff}=2c_2\tilde{\alpha}^4 \bigg[\frac{1}{4}{\phi}^4-\frac{1}{3}{\phi}^3+\frac{1}{4}m^2({\phi}^2-1)^2\bigg]+4c_2\log{\phi}\label{formula}
\end{eqnarray}
The equation of motion $\partial V_{\rm eff}/ \partial \phi =0$ admits two real solutions where we can identify the one with the least energy  with the actual radius of the sphere. However this is only true up to a certain value 
$\tilde{\alpha}_{*}$ of the coupling constant $\tilde{\alpha}$ where no real solution will exist and as a consequence the fuzzy sphere solution $X_a=\alpha {\phi}L_a$ will not exist. In other words the potential $V_{\rm eff}$ below the value $\tilde{\alpha}_{*}$ becomes unbounded and the fuzzy sphere  collapses. The critical values can be easily computed and one finds by extrapolating to large masses  $
{\phi}_{*}={1}/{\sqrt{2}}$
and
\begin{eqnarray}
\tilde{\alpha}_{*}=\big[\frac{8}{m^2+\sqrt{2}-1}\big]^{\frac{1}{4}}. \label{pre2}
\end{eqnarray}
In other words the phase transition happens each time at a smaller value of the coupling constant $\tilde{\alpha}$ and thus the fuzzy sphere is more stable. The critical value $\tilde{\alpha}_{*}$ separates the "fuzzy sphere phase" where we have a $U(1)$ gauge theory on the fuzzy sphere ${\bf S}^2_N$ from the  "matrix phase" where this picture breakes down completely.

The UV-IR mixing is seen at this non-perturbative level as a transition between completely different phases of the theory. Indeed by crossing to 
 the matrix phase the radius of the sphere goes to zero and hence the noncommutativity parameter which is proportional to $R$  in the planar limit will also go to zero. This  is the singlar limit of the UV-IR mixing discussed above.  The fact that (\ref{pre2}) approaches zero when $m{\longrightarrow}\infty $ means that reaching zero radius becomes more difficult as we increase $m$ and as a consequence the singular limit $\theta {\longrightarrow}0$ becomes also harder to reach ( i.e smooth ) for these large values of $m$. Thus from one hand the fuzzy sphere is becoming more stable and the matrix phase is shrinking while from the other hand the UV-IR mixing is becoming vansihingly small as $m{\longrightarrow}\infty$ which is our main observation that the two effects must be related at least in this case. 

The perturbative UV-IR mixing is a typical property of quantum field theories on noncommutative spaces  which derives from the noncommutativity with no commutative  analogue \cite{others1,min,szabor}. At the non-perturbative level this mixing may be related to topology change. In this case the  $2$ dimensional spacetime ( the fuzzy sphere ) collapses onto a point ( the matrix phase ) under quantum effects. The UV-IR mixing in this picture is ( possibly ) a reflection of the fact that spacetime itself may evaporates when quantum fluctuations of fields are taken into consideration in the presence of a non-zero noncommutativity. The noncommutativity somehow made it possible that fields and spacetime talk to each other in the same way that gravity does. A very concrete way in implementing this scenario are  the matrix models (\ref{main2}) and their generalization to other fuzzy spaces such as fuzzy ${\bf CP}^2$ \cite{in-progress}. See \cite{CP2} for other discussions of fuzzy ${\bf CP}^2$, \cite{denjoe10} for higher fuzzy  ${\bf CP}^n$ and \cite{S4} for fuzzy ${\bf S}^4$. We believe that most fuzzy spaces will have topology change in the same way that most noncommutative Moyal-Weyl spaces will have UV-IR mixing.

However the action (\ref{main2}) does not know a priori about all the above perturbative and semi-non-perturbative statements which rely on our choice of the vacuum  (\ref{fuzz}) and on the different scaling limits considered.  So the nonperturbative behaviour of the model for small values of $\alpha $ ( or equivalently  large values of $g$ ) is not obvious. A fully nonperturbative study of the $U(1)$ model is done by using Monte Carlo simulations with the Metropolis algorithm and the action (\ref{main2}) in \cite{ref1}. In particular we compute the phase diagram of the model. 
In  \cite{nishimura} the study was done for $m=0$.

There are three different phases of  $U(1)$ gauge theory on ${\bf S}^2_N$. In the "matrix phase" the fuzzy sphere vacuum (\ref{fuzz}) collapses under quantum fluctuations and there is  no underlying sphere in the continuum large $N$ limit or underlying Moyal-Weyl plane in the noncommutative planar limit. This is expected from perturbation theory and the effective potential calculation. In this phase we have instead a $U(N)$  theory on a point.

The other phase is the ``fuzzy sphere phase'' where (\ref{fuzz}) is stable. We  observe that the fuzzy sphere phase splits into two distinct regions corresponding to the weak and strong coupling phases of the gauge field. These are separated by a third order phase transition which is consistent with that of a  one-plaquette  model \cite{in-progress,ref10}. This was not detected in perturbation theory. The gauge field in this phase ( in particular across the critical line and inside the strong coupling phase ) behaves as if it is a large $U(N)$ commutative gauge theory on a lattice. Although classically and in the very weak coupling phase the model is a $U(1)$ on ${\bf S}^2_N$.  This $U(N)$ behaviour in the limit is consistent with the fact that we have $U(N)$ in the matrix phase. So the effect of the matrix  phase on the structure of the gauge group survives  even after we cross to the fuzzy sphere phase. 
However in the light of the above perturbative calculations there is still a possibility that the model (\ref{main2}) with $m$ fixed to some power of $N$ ( so it is not a free parameter anymore) will not show this one-plaquette critical line \cite{in-progress}. This is also the expectation of  \cite{steinackers2}.

\section{Classical considerations on fuzzy ${\bf S}^2\times {\bf S}^2$}

$U(1)$ gauge field on fuzzy ${\bf S}^2{\times}{\bf S}^2$ is associated with a set of six hermitian $(L+1)^2\times (L+1)^2$ matrices $D_{AB}$ ($D_{AB}=-D_{BA}$, $A,B=1,4$) which transform homogeneously under the action of the group, i.e
\begin{equation}
D_{AB}\to UD_{AB}U^{-1}, ~~~~U\in U\left((L+1)^2\right).
\end{equation}
The action is given by (with $Tr_{L}=\frac{1}{\left(L+1\right)^{2}}Tr$, $Tr_L{\bf 1}_{(L+1)^2}=(L+1)^2$, $g$ is the gauge coupling constant and $m$ is the mass of the normal components of the gauge field )

\begin{eqnarray}
S& = & \frac{1}{16g^2}\left\{-\frac{1}{4}Tr_{L}[D_{AB},D_{CD}]^2+\frac{i}{3}f_{ABCDEF}Tr_{L}[D_{AB},D_{CD}]D_{EF}\right\}\nonumber\\
  &   &  \nonumber \\
  & + &  \frac{m^2}{8g^2L^2_{AB}}Tr_{L}(D^2_{AB}-L^2_{AB})^2+\frac{m^2}{32g^2L^2_{AB}}Tr_{L}(\epsilon_{ABCD}D_{AB}D_{CD})^2.
\label{model}
\end{eqnarray}
In the above action $f_{ABCDEF}$ are the structure constants of the Lie algebra $so(4)$. Indeed the generators $L_{AB}$ ( with $L_{AB}=-L_{BA}$ ) satisfy the commutation relations
\begin{eqnarray}
[L_{AB},L_{CD}]=if_{ABCDEF}L_{EF}&=&\frac{i}{2}\bigg({\delta}_{BC}L_{AD}-{\delta}_{BD}L_{AC}+{\delta}_{AD}L_{BC}-{\delta}_{AC}L_{BD}\bigg)\nonumber\\
&-&\frac{i}{2}\bigg({\delta}_{DA}L_{CB}-{\delta}_{DB}L_{CA}+{\delta}_{CB}L_{DA}-{\delta}_{CA}L_{DB}\bigg).
\end{eqnarray}
The equations of motion are given by 
\begin{eqnarray}
i[D_{CD},F_{AB,CD}]+
\frac{4m^2}{\sqrt{c_2}}\{D_{AB},{\Phi}_1+{\Phi}_2\}+\frac{m^2}{\sqrt{c_2}}\{\epsilon_{ABCD}D_{CD},{\Phi}_1-{\Phi}_2\} =0. 
\end{eqnarray}
In above the $SU(2)$ Casimir $c_2$ is given by $c_2=\frac{L}{2}(\frac{L}{2}+1)$. As we will show $F_{AB,CD}=i\left[D_{AB},D_{CD}\right]+f_{ABCDEF}D_{EF}$ is the curvature of the gauge field on fuzzy ${\bf S}^{2}\times {\bf S}^{2}$ whereas ${\Phi}_1$ and ${\Phi}_2$ (defined by $D_{AB}^2-L_{AB}^2=8\sqrt{c_2}({\Phi}_1+{\Phi}_2)$ and ${\epsilon}_{ABCD}D_{AB}D_{CD}=16\sqrt{c_2}({\Phi}_1-{\Phi}_2)$) are  the  normal components of the gauge field on ${\bf S}^{2}\times {\bf S}^{2}$.

The most obvious non-trivial solution of the equations of motion must satisfy $F_{AB,CD}=0$, $D^{2}_{AB}=L^{2}_{AB}$ and $\epsilon_{ABCD}D_{AB}D_{CD}=0$ (or equivalently $F_{AB}=0$, ${\Phi}_i=0$). This solution is clearly given by the generators $L_{AB}$ of the irreducible representation $(\frac{L}{2},\frac{L}{2})$  of $SO(4)$, viz
\begin{eqnarray}
D_{AB}=L_{AB}.\label{sol}
\end{eqnarray}
By expanding $D_{AB}$ around this vacuum as $D_{AB}=L_{AB}+A_{AB}$ and substituting back into the action (\ref{model}) we obtain a $U(1)$ gauge field $A_{AB}$ on ${\bf S}^2_L{\times}{\bf S}^2_L$. The matrices $D_{AB}$ are thus  the covariant derivatives on ${\bf S}^2_L{\times}{\bf S}^2_L$. The curvature $F_{AB,CD}$ in terms of $A_{AB}$ takes the usual form $F_{AB,CD}=i{\cal L}_{AB}A_{CD}-i{\cal L}_{CD}A_{AB}+{f}_{ABCDEF}A_{EF}+i[A_{AB},A_{CD}]$. The normal scalar fields in terms of $A_{AB}$ are on the other hand given by $8\sqrt{c_2}({\Phi}_1+{\Phi}_2)=L_{AB}A_{AB}+A_{AB}L_{AB}+A_{AB}^2$ and $16\sqrt{c_2}({\Phi}_1-{\Phi}_2)={\epsilon}_{ABCD}(L_{AB}A_{CD}+A_{AB}L_{CD}+A_{AB}A_{CD})$. 

The true gauge field on fuzzy ${\bf S}^{2}\times {\bf S}^{2}$ must in fact be $4-$dimensional ( as opposed to $A_{AB}$ which is $6-$dimensional ) and hence the extra two components considered in this description are scalar fields which are the normal components of $A_{AB}$ on ${\bf S}^{2}\times {\bf S}^{2}$. In the fuzzy setting there is no known  covariant splitting of $A_{AB}$ into a  $4-$dimensional tangent gauge field and the above normal components .

In order to discuss the continuum limit of the action (\ref{model}) we  introduce the  matrices $D^{(1)}_{a}=L_a^{(1)}+A_a^{(1)}$ and $ D^{(2)}_{a}=L_a^{(2)}+A_a^{(2)}$ defined by 
\begin{eqnarray}
D^{(1)}_{a}{\equiv}-\frac{1}{2}\left[\frac{1}{2}\epsilon_{abc}D_{bc}+D_{a4}\right]~,~D^{(2)}_{a}{\equiv}-\frac{1}{2}\left[\frac{1}{2}\epsilon_{abc}D_{bc}-D_{a4}\right].\label{la} 
\end{eqnarray}
Clearly $D_a^{(1)}$ ($A_a^{(1)}$)  and $D_a^{(2)}$ ($A_a^{(2)}$) are the components of $D_{AB}$ ($A_{AB}$) on the two spheres respectively. The curvature becomes $F^{(ij)}_{ab}=i\mathcal{L}^{(i)}_{a}A^{(j)}_{b}-i\mathcal{L}^{(j)}_{b}A^{(i)}_{a}+{\delta}_{ij}\epsilon_{abc}A^{(i)}_{c}+i[A_a^{(i)},A_b^{(j)}]$ whereas the normal scalar fields become $2\sqrt{c_2}{\Phi}_i=(D_a^{(i)})^2-c_2=L_a^{(i)}A_a^{(i)}+A_a^{(i)}L_a^{(i)}+(A_a^{(i)})^2$. In terms of this three dimensional notation the action (\ref{model}) reads 
\begin{eqnarray}
&&S=S^{(1)}+S^{(2)}+S^{(1,2)}\nonumber\\
&&S^{(1,2)}=\frac{1}{2g^2}Tr_{L}\left(F_{ab}^{(12)}\right)^2.\label{ac}
\end{eqnarray}
$S^{(1)}$ and $S^{(2)}$ are the actions for the $U(1)$ gauge fields $A_a^{(1)}$ and $A_a^{(2)}$ on a single fuzzy sphere ${\bf S}^2_L$. They are given by
\begin{eqnarray}
S^{(i)}
&=&\frac{1}{4g^2}Tr_{L}\left(F_{ab}^{(i)}\right)^2-\frac{1}{2g^2}{\epsilon}_{abc}Tr_L\left[\frac{1}{2}F_{ab}^{(i)}A_c^{(i)}-\frac{i}{6}[A_a^{(i)},A_b^{(i)}]A_c^{(i)}\right]+\frac{2m^2}{g^2}Tr_L{\Phi}_i^2.\nonumber\\&&
\end{eqnarray}
It is immediately clear that in the continuum limit  $L{\longrightarrow}{\infty}$ the action (\ref{ac}) describes the interaction of a genuine $4-$d gauge field with the normal scalar fields ${\Phi}_i=n_a^{(i)}A_a^{(i)}$ where $n_a^{(i)}$ is the unit normal vector to the $i$-th sphere. Let us also remark that in this limit the $3-$dimensional fields $A_a^{(i)}$ decompose as $A_a^{(i)}=(A_a^{(i)})^T+n_a^{(i)}{\Phi}_i$ where $(A_a^{(i)})^T$ are the tangent $2-$dimensional gauge fields. Since the differential calculus on ${\bf S}^2 \times {\bf S}^2$ is intrinsically $6-$dimensional we can not decompose the fuzzy gauge field in a similar (gauge-covariant) fashion and as a consequence we can not write an action on the fuzzy ${\bf S}^2 \times {\bf S}^2$ which will only involve the desired $4-$dimensional gauge field. 

\section{Note on Monte Carlo simulations and matrix models}
Before we proceed to the one-loop quantum theory let us say few words about Monte Carlo simulations of the above model. The action $S^{(1)}$ on the first sphere can be put in the form
\begin{eqnarray}
S^{(1)}&=&N\bigg[-\frac{1}{4}Tr[X_a,X_b]^2+\frac{2i{\alpha}}{3}{\epsilon}_{abc}TrX_aX_bX_c\bigg]-Nm^2{\alpha}^2 TrX_a^2+\frac{Nm^2}{2c_2} Tr(X_a^2)^2.
\end{eqnarray}
The action $S^{(2)}$ on the second sphere is similarly given by
\begin{eqnarray}
S^{(2)}&=&N\bigg[-\frac{1}{4}Tr[Y_a,Y_b]^2+\frac{2i{\alpha}}{3}{\epsilon}_{abc}TrY_aY_bY_c\bigg]-Nm^2{\alpha}^2 TrY_a^2+\frac{Nm^2}{2c_2} Tr(Y_a^2)^2.
\end{eqnarray}
Now $N=(L+1)^2$, $X_a=\alpha D_a^{(1)}$, $Y_a=\alpha D_a^{(2)}$, $N{\alpha}^4=\tilde{\alpha}^4/N=1/(Ng^2)$. The coupling between the two spheres is given by the action
\begin{eqnarray}
S^{(12)}=-\frac{N}{2}Tr[X_a,Y_b]^2.
\end{eqnarray}
The action which will be relevant for the numerical simulation is the matrix model given by the sum $S^{(1)}+S^{(2)}+S^{(12)}$ \cite{in-progress}.

To study the noncommutative planar limit we should instead consider the following action on the first sphere \cite{in-progress}
\begin{eqnarray}
S^{(1)}&=&NTr\bigg[-\frac{1}{4}[X_a,X_b]^2+i{\alpha}{\epsilon}_{abc}X_aX_bX_c+\frac{{\alpha}^2}{2}X_a^2\bigg]-Nm^2{\alpha}^2 TrX_3^2+\frac{Nm^2}{2c_2} Tr(X_3^2)^2 .\label{model1}\nonumber\\
\end{eqnarray}
In above $X_a=\alpha R D_a^{(1)}$, $g^2=1/(N^2R^2{\alpha}^4)$ and $m=N^p$ with some power $p$. The first term is the action (\ref{action2}) without the Chern-Simons-like terms and without the mass term whereas the second term implements  in the limit the constraint $D_3=R/{\theta}^2$  which means that we are restricted to the north pole in a covariant way. This action is gauge invariant but not rotationally invariant. For the seond sphere we should write a similar action whereas the coupling between the two spheres remains unchanged.

\section{The one-loop quantum effective action}
The partition function of the theory depends  on $3$
parameters, the Yang-Mills coupling constant $g$, the mass $m$ of the normal scalar fields, and the size $L$ of the matrices.
Using the background field method 
we obtain the one-loop effective action
\begin{equation}
\Gamma\left[D_{AB}\right]=S\left[D_{AB}\right]+\frac{1}{2}Tr_{6}TR\log{\Omega}_{ABCD}-TR\log\mathcal{D}_{AB}^{2}.
\label{effaction}
\end{equation}
${\Omega}_{ABCD}$ is defined by
\begin{eqnarray}
{\Omega}_{ABCD}&=&\frac{1}{2}\mathcal{D}_{EF}^2{\delta}_{AB,CD}-\left(1-\frac{1}{\xi}\right)\mathcal{D}_{AB}\mathcal{D}_{CD}-2i\mathcal{F}_{ABCD}+\frac{4 m^2}{L^{2}_{AB}}\Omega^{\left(1\right)}_{ABCD},
\end{eqnarray}
where $\delta_{AB,CD}=\delta_{AC}\delta_{BD}-\delta_{AD}\delta_{BC}$, and 
\begin{eqnarray}
\Omega^{\left(1\right)}_{ABCD}&=&(D^2_{EF}-L^2_{EF})\delta_{AB,CD}+\frac{1}{2}(\epsilon_{EFGH}D_{EF}D_{GH})\epsilon_{ABCD}\nonumber\\
&&-\mathcal{D}_{AB}\mathcal{D}_{CD}
-\widetilde{\mathcal{D}}_{AB}\widetilde{\mathcal{D}}_{CD}
+4D_{AB}D_{CD}+4\widetilde{D}_{AB}\widetilde{D}_{CD}.
\end{eqnarray}
The notation ${\mathcal D}_{AB}$ and ${\mathcal F}_{ABCD}$ means that
the covariant derivative $D_{AB}$ and the curvature $F_{ABCD}$ act by
commutators, i.e ${\cal D}_{AB}(M)=[D_{AB},M]$, ${\cal
F}_{ABCD}(M)=[F_{ABCD},M]$ where $M$ is an element of
$Mat_{(L+1)^{2}}$. We have also introduced the notation $\widetilde{D}_{AB}\equiv\frac{1}{2}\epsilon_{ABCD}D_{CD}$. $TR$ is the trace over the $4$ indices corresponding to the left and  right actions of  operators on matrices. $Tr_{6}$ is the trace associated with the action of $SU(2){\times}SU(2)$.

In the remainder of this letter we will use mainly three dimensional indices. The effective action simplifies considerably in the Feynman gauge $\xi=1$ and for  $m=0$. We can compute
\begin{eqnarray}
\frac{1}{2}Tr_{6}TR\log{\Omega}_{ABCD}&=&\int dX_a^{(1)}e^{-4TrX_a^{(1)}{\Omega}_{ab}^{(1)}X_b^{(1)}}\nonumber\\
&{\times}&\int dX_a^{(2)}e^{-4TrX_a^{(2)}{\Omega}_{ab}^{(2)}X_b^{(2)}}e^{32iTr(X_a^{(1)}{\cal F}_{ab}^{(12)}X_b^{(2)}+X_a^{(2)}{\cal F}_{ab}^{(21)}X_b^{(1)})}\nonumber\\
&=&\int dX_a^{(1)}e^{-4TrX_a^{(1)}\big({\Omega}_{ab}^{(1)}+{\Omega}_{ab}^{(12)}\big)X_b^{(1)}}\nonumber\\
&{\times}&\int dX_a^{(2)}e^{-4Tr\big(X^{(2)}-8iX^{(1)}{\cal F}^{(12)}{\Omega}^{(2)-1}\big)_a{\Omega}_{ab}^{(2)}\big(X^{(2)}-8i{\Omega}^{(2)-1}{\cal F}^{(21)}X^{(1)}\big)_b}.
\end{eqnarray}
where ${\Omega}_{ab}^{(i)}={\cal D}_{AB}^2{\delta}_{ab}-8i{\cal F}_{ab}^{(i)}$, ${\Omega}^{(12)}=64{\cal F}^{(12)}{\Omega}^{(2)-1}{\cal F}^{(21)}$ and ${F}_{ab}^{(i)}=i[D_a^{(i)},D_b^{(i)}]+{\epsilon}_{abc}D_c^{(i)}$, ${F}_{ab}^{(ij)}=i[D_a^{(i)},D_b^{(j)}]$. In above we have also used the identity $X_{AB}{\cal O}Y_{AB}=4X_a^{(1)}{\cal O}Y_{a}^{(1)}+4X_a^{(2)}{\cal O}Y_{a}^{(2)}$ and the identity $f_{ABCDEF}TrX_{AB}Y_{CD}Z_{EF}=16{\epsilon}_{abc}TrX_a^{(1)}Y_b^{(1)}Z_c^{(1)}+16{\epsilon}_{abc}TrX_a^{(2)}Y_b^{(2)}Z_c^{(2)}$. Hence by using the three dimensional notation the effective action takes the form

\begin{eqnarray}\label{effective}
\Gamma (D)=S(D)+\frac{1}{2}Tr_3
TR \log \left(\Omega^{(1)}_{ab}+{\Omega}^{(12)}_{ab}\right)+\frac{1}{2}Tr_3 TR \log \Omega^{(2)}_{ab}-TR
\log \mathcal{D}^2_{AB}.
\end{eqnarray}
The quadratic effective action is obtained by keeping powers up to $2$ in the gauge field. We obtain the action
\begin{eqnarray}
\Gamma_2^{} &=&\Gamma_2^{(1)}+ \Gamma_2^{(2)}+\Gamma_2^{(1,2)}
\end{eqnarray}
where
\begin{eqnarray}
\Gamma_2^{(i)}&=&S_2^{(i)}+2TR\frac{1}{\Delta}\bigg(\mathcal{L}_{a}^{(i)}\mathcal{A}_{a}^{(i)}+\mathcal{A}_{a}^{(i)}\mathcal{L}_{a}^{(i)}+({\cal A}^{(i)}_a)^2 \bigg)\nonumber\\
&-&TR\frac{1}{\Delta}\bigg(\mathcal{L}_{a}^{(i)}\mathcal{A}_{a}^{(i)}+\mathcal{A}_{a}^{(i)}\mathcal{L}_{a}^{(i)}\bigg)\frac{1}{\Delta}\bigg(\mathcal{L}_{a}^{(i)}\mathcal{A}_{a}^{(i)}+\mathcal{A}_{a}^{(i)}\mathcal{L}_{a}^{(i)}\bigg)-TR\frac{1}{\Delta}\mathcal{F}^{(i)}_{ab} \frac{1}{\Delta}\mathcal{F}^{(i)}_{ab}  \nonumber\\
 \Gamma_2^{(1,2)} &=&S_2^{(1,2)}-2TR\frac{1}{\Delta}\bigg(\mathcal{L}_{a}^{(1)}\mathcal{A}_{a}^{(1)}+\mathcal{A}_{a}^{(1)}\mathcal{L}_{a}^{(1)}\bigg)\frac{1}{\Delta}\bigg(\mathcal{L}_{a}^{(2)}\mathcal{A}_{a}^{(2)}+\mathcal{A}_{a}^{(2)}\mathcal{L}_{a}^{(2)}\bigg)-2TR\frac{1}{\Delta}\mathcal{F}^{(12)}_{ab} \frac{1}{\Delta}\mathcal{F}^{(21)}_{ab}  .\nonumber\\\label{dlta}
\end{eqnarray}
The Laplacian ${\Delta}$ is defined by 
\begin{eqnarray}
{\Delta}=({\cal L}_a^{(1)})^2+({\cal L}_a^{(2)})^2. 
\end{eqnarray}
As before ${\cal L}_a^{(i)}$ and ${\cal A}_a^{(i)}$ act by commutators $[L_a^{(i)},..]$ and $[A_a^{(i)},..]$ and $L_a^{(i)}$ and $A_a^{(i)}$ are defined in terms of $L_{AB}$ and $A_{AB}$ respectively by equations similar to (\ref{la}). Furthermore $S_2^{(i)}$ and $S_2^{(1,2)}$ are the quadratic parts of the classical actions $S^{(i)}$ and $S^{(1,2)}$ respectively.

In the following we will consider without any loss of generality the background configuration in which $A_a^{(2)}=0$. In other words we will study the background matrices 
\begin{eqnarray}
D_a^{(1)}=L_a^{(1)}+A_a^{(a)}~,~ D_a^{(2)}=L_a^{(2)}. \label{16}
\end{eqnarray}
If it can be shown that there exists a UV-IR mixing phenomena in this case then we should conclude immediately that there must exist a UV-IR mixing phenomena in the general case since extension to the case  $A_a^{(2)}{\neq}0$ is rather straightforward and trivial. The effective action reads for this configuration
\begin{eqnarray}
\Gamma_2^{} &=&\Gamma_2^{(1)}+\Gamma_2^{(1,2)}\nonumber\\
&=&S_2^{(1)}+S_2^{(1,2)}+2TR\frac{1}{\Delta}\bigg(\mathcal{L}_{a}^{(1)}\mathcal{A}_{a}^{(1)}+\mathcal{A}_{a}^{(1)}\mathcal{L}_{a}^{(1)}+({\cal A}^{(1)}_a)^2 \bigg)\nonumber\\
&-&TR\frac{1}{\Delta}\bigg(\mathcal{L}_{a}^{(1)}\mathcal{A}_{a}^{(1)}+\mathcal{A}_{a}^{(1)}\mathcal{L}_{a}^{(1)}\bigg)\frac{1}{\Delta}\bigg(\mathcal{L}_{a}^{(1)}\mathcal{A}_{a}^{(1)}+\mathcal{A}_{a}^{(1)}\mathcal{L}_{a}^{(1)}\bigg)-TR\frac{1}{\Delta}\mathcal{F}^{(1)}_{ab} \frac{1}{\Delta}\mathcal{F}^{(1)}_{ab}-2TR\frac{1}{\Delta}\mathcal{F}^{(12)}_{ab} \frac{1}{\Delta}\mathcal{F}^{(12)}_{ab}  .\nonumber\\
\end{eqnarray}
Remark in particular that $F_{ab}^{(12)}=-i[L_b^{(2)},A_a^{(1)}]$. Any function on fuzzy ${\bf S}^2{\times}{\bf S}^2$ can be expanded in
terms of the basis

\begin{eqnarray}
\hat{\mathcal{Y}}_{l_1m_1;l_2m_2}=\hat{Y}_{l_1m_1}\otimes\hat{Y}_{l_2m_2}.
\end{eqnarray}
$\hat{Y}_{lm}$ are the standard $SU(2)$ polarization tensor \cite{VKM}. For example the gauge field $A_a^{(1)}$ is expanded as
\begin{eqnarray}
A_a^{(1)}=\sum_{l_1m_1,l_2m_2}A_a(l_1m_1,l_2m_2)\hat{\mathcal{Y}}_{l_1m_1;l_2m_2}.
\end{eqnarray}
The 2-point Green's function is given by

\begin{eqnarray}
\left(\frac{1}{\Delta}\right)^{AB,CD}=\frac{1}{(L+1)^2}\sum_{l_1m_1}\sum_{l_2m_2}\frac{(\yy_{l_1m_1;l_2m_2})^{AB}(\yyd_{l_1m_1;l_2m_2})^{DC}}{[l_1(l_1+1)+l_2(l_2+1)]}
\end{eqnarray}
In this formula the $A$,$B$,$C$ and $D$ are matrix indices ( and not $SO(4)$ indices ) so they run over the range $1,...,(L+1)^2$. 

The Tadpole contribution is given by
\begin{eqnarray}
2TR\frac{1}{\Delta}\bigg(\mathcal{L}_{a}^{(1)}\mathcal{A}_{a}^{(1)}+\mathcal{A}_{a}^{(1)}\mathcal{L}_{a}^{(1)} \bigg)&=&-4\sum_{l_1m_1,l_2m_2}\frac{Tr_L[L_a^{(1)},\hat{\mathcal{Y}}_{l_1m_1;l_2m_2}^+][A_a^{(1)},\hat{\mathcal{Y}}_{l_1m_1;l_2m_2}]}{l_1(l_1+1)+l_2(l_2+1)}\nonumber\\
&=&-4\sum_{k_1n_1,k_2n_2}A_a^{(1)}(k_1n_1,k_2n_2){\gamma}_a(k_1n_1,k_2n_2),
\end{eqnarray}
where
\begin{eqnarray}
{\gamma}_a(k_1n_1,k_2n_2)=\sum_{l_1m_1,l_2m_2}\frac{Tr_L[L_a^{(1)},\hat{\mathcal{Y}}_{l_1m_1;l_2m_2}^+][\hat{\mathcal{Y}}_{k_1n_1;k_2n_2},\hat{\mathcal{Y}}_{l_1m_1;l_2m_2}]}{l_1(l_1+1)+l_2(l_2+1)}.
\end{eqnarray}
A short calculation ( see the appendix ) yields the result
\begin{eqnarray}
2TR\frac{1}{\Delta}\big(\mathcal{L}_{a}^{(1)}\mathcal{A}_{a}^{(1)}+\mathcal{A}_{a}^{(1)}\mathcal{L}_{a}^{(1)} \big)&=&8\hat{c}_2\sqrt{c_2}Tr_L{\Phi}_1-4\hat{c}_2Tr_L(A_a^{(1)})^2.
\end{eqnarray}
In above $\hat{c}_2$ is given by
\begin{eqnarray}
\hat{c}_2&=&\frac{2}{L(L+2)}\sum_{k_1=1}\sum_{k_2=0}\frac{(2k_1+1)(2k_2+1)}{k_1(k_1+1)+k_2(k_2+1)}k_1(k_1+1)\label{c2}
\end{eqnarray}
or equivalently $\hat{c}_2=\frac{(L+1)^4-1}{L(L+2)}=L^2+2L+2$. 

The vacuum polarization diagrams are also computed in the appendix. Let us summarize the results.
The $4-$vertex contribution is

\begin{eqnarray}
2TR\frac{1}{\Delta}\bigg(\mathcal{A}_{a}^{(1)} \bigg)^2&=&-2\sum_{l_1m_1,l_2m_2}\frac{Tr_L[A_a^{(1)},\hat{\mathcal{Y}}_{l_1m_1;l_2m_2}^+][A_a^{(1)},\hat{\mathcal{Y}}_{l_1m_1;l_2m_2}]}{l_1(l_1+1)+l_2(l_2+1)}.
\end{eqnarray}
This  can be  computed quite easily and one  finds the result
\begin{eqnarray}
2TR\frac{1}{\Delta}\bigg(\mathcal{A}_{a}^{(1)} \bigg)^2&=&Tr_LA_a^{(1)}{\cal O}_4({\Delta}_1,{\Delta }_2)A_a^{(1)}.
\end{eqnarray}
Clearly ${\Delta}_1=({\cal L}_a^{(1)})^2$ and ${\Delta}_2=({\cal L}_a^{(2)})^2$ are the Laplacians on the two spheres separately. The opeartor ${\cal O}_4$ is defined by its eigenvalues ${\cal O}_4(p_1,p_2)$ ( given in equation (\ref{39}) ) on its eigenvectors $\hat{\mathcal{Y}}_{p_1s_1;p_2s_2}$.

Similarly the $F-$vertex contribution can be computed and one finds 

\begin{eqnarray}
-TR\frac{1}{{\Delta}}\mathcal{F}^{(1)}_{ab}\frac{1}{{\Delta}}\mathcal{F}^{(1)}_{ab}&=&-Tr_LF_{ab}^{(1)}{\cal O}_F({\Delta}_1,{\Delta }_2)F_{ab}^{(1)}.
\end{eqnarray}
The opeartor ${\cal O}_F$ is defined by the eigenvalues ${\cal O}_F(p_1,p_2)$ ( given in equation (\ref{46}) ) on  the eigenvectors $\hat{\mathcal{Y}}_{p_1s_1;p_2s_2}$. By analogy we will have

\begin{eqnarray}
-2TR\frac{1}{{\Delta}}\mathcal{F}^{(12)}_{ab}\frac{1}{{\Delta}}\mathcal{F}^{(12)}_{ab}&=&-2Tr_LF_{ab}^{(12)}{\cal O}_F({\Delta}_1,{\Delta }_2)F_{ab}^{(12)}\nonumber\\
\end{eqnarray}
Finally we need to compute the $3-$vertex correction
\begin{eqnarray}
-TR\frac{1}{\Delta}\bigg(\mathcal{L}_{a}^{(1)}\mathcal{A}_{a}^{(1)}+\mathcal{A}_{a}^{(1)}\mathcal{L}_{a}^{(1)}\bigg)\frac{1}{\Delta}\bigg(\mathcal{L}_{a}^{(1)}\mathcal{A}_{a}^{(1)}+\mathcal{A}_{a}^{(1)}\mathcal{L}_{a}^{(1)}\bigg)\label{27}
\end{eqnarray}
This is by far the most difficult calculation. In the last part of the appendix  we find that this correction gives two different contribution to the effective action. The most important is a canonical gauge contribution of the form
\begin{eqnarray}
Tr_L{\cal L}_a^{(1)}A_a^{(1)}{\cal O}_3({\Delta}_1,{\Delta}_2){\cal L}_b^{(1)}A_b^{(1)}
\end{eqnarray}
As before the operator ${\cal O}_3({\Delta}_1,{\Delta}_2)$ is defined by its eigenvalues ${\cal O}_3(p_1,p_2)$ ( given in equation (\ref{61}) ) on its eigenvectors $\hat{\mathcal{Y}}_{p_1s_1;p_2s_2}$.  

The other contribution in the $3-$vertex correction (\ref{27}) is of scalar-type ( in other words it involves anticommutators between $A_a$ and $L_a$ instead of commutators ) and it was studied in detail in \cite{ref}. See also the appendix.

Putting all the above results together we obtain ( modulo scalar-type terms ) the full effective quadratic action in the form
\begin{eqnarray}
\Gamma_2&=&S_2^{(1)}+S_2^{(1,2)}-Tr_LF_{ab}^{(1)}{\cal O}_FF_{ab}^{(1)}+Tr_L{\cal L}_a^{(1)}A_a^{(1)}{\cal O}_3{\cal L}_b^{(1)}A_b^{(1)}-2Tr_LF_{ab}^{(12)}{\cal O}_F F_{ab}^{(12)}\nonumber\\
&+&Tr_LA_a^{(1)}\big[{\cal O}_4-4\hat{c}_2\big]A_a^{(1)}.
\end{eqnarray}
We use the identity
\begin{eqnarray}
Tr_L{\cal L}_a^{(1)}A_a^{(1)}{\cal O}_3{\cal L}_b^{(1)}A_b^{(1)}&=&\frac{1}{2}Tr_LF_{ab}^{(1)}{\cal O}_3F_{ab}^{(1)}-\frac{1}{2}{\epsilon}_{abc}Tr_LF_{ab}^{(1)}{\cal O}_3A_a^{(1)}-Tr_LA_a^{(1)}{\cal O}_3{\Delta}_1A_a^{(1)}
\end{eqnarray}
Hence
\begin{eqnarray}
\Gamma_2&=&S_2^{(1)}+S_2^{(1,2)}+Tr_LF_{ab}^{(1)}(\frac{1}{2}{\cal O}_3-{\cal O}_F)F_{ab}^{(1)}-\frac{1}{2}{\epsilon}_{abc}Tr_LF_{ab}^{(1)}{\cal O}_3A_a^{(1)}\nonumber\\
&-&2Tr_LF_{ab}^{(12)}{\cal O}_F F_{ab}^{(12)}+Tr_LA_a^{(1)}\big[{\cal O}_4-{\cal O}_3{\Delta}_1-4\hat{c}_2\big]A_a^{(1)}.\label{32}
\end{eqnarray}
The eigenvalues of the operators  ${\cal O}_3$,${\cal O}_4$ and ${\cal O}_F$   are given from the results of the appendix by 
\begin{eqnarray}
{\cal O}_i(k_1l_1;p_1p_2)&=&4(L+1)^2\sum_{k_1,k_2}\sum_{l_1,l_2}\frac{(2k_1+1)(2k_2+1)}{k_1(k_1+1)+k_2(k_2+1)}\frac{(2l_1+1)(2l_2+1)}{l_1(l_1+1)+l_2(l_2+1)}[1-(-1)^{R+p_1+p_2}]\nonumber\\
&\times &\left\{\begin{array}{ccc}
                   p_1 & k_1 & l_1 \\
            \frac{L}{2} & \frac{L}{2}  & \frac{L}{2}
                 \end{array}\right\}^2\left\{\begin{array}{ccc}
                   p_2 & k_2 & l_2 \\
            \frac{L}{2} & \frac{L}{2}  & \frac{L}{2}
                 \end{array}\right\}^2
x_i(k_1k_2l_1l_2;p_1p_2)\label{6j}
\end{eqnarray}
where
\begin{eqnarray}
x_3(k_1k_2l_1l_2;p_1p_2)&=&-\frac{k_1(k_1+1)\big(l_1(l_1+1)-k_1(k_1+1)\big)}{p_1^2(p_1+1)^2}\nonumber\\
x_4(k_1k_2l_1l_2;p_1p_2)&=&k_1(k_1+1)+k_2(k_2+1)\nonumber\\
x_F(k_1k_2l_1l_2;p_1p_2)&=&\frac{1}{2}.\label{6j+}
\end{eqnarray}
In the loop integrals ${\cal O}_i$ the $1$ in $1-(-1)^{R+p_1+p_2}$ corresponds to planar diagrams while the $(-1)^{R+p_1+p_2}$ corresponds to non-planar diagrams as we will explain below. The quantum number $R$ ( not to be confused with the radius of the sphere ) is given by $R=k_1+k_2+l_1+l_2$. The pair $(p_1,p_2)$ represents the external momentum. The pairs $(k_1,k_2)$ and $(l_1,l_2)$ represent internal momenta. The factors $2j+1$ give the volume forms ( similar to $dp$ on the plane ) whereas $1/(j_1(j_1+1)+j_2(j_2+1))$ give propagators  similar to $1/p^2$ on the $4-$dimensional ${\bf R}^4$. The $6j$ symbols encode energy conservation rules. The complicated interactions of the fuzzy photon are reflected in the $6j$ symbols and the coefficients $x_i$.

\section{The UV-IR mixing in the planar limit ${\bf R}^2\times {\bf R}^2_{\theta}$}

The analysis of the quadratic effective action (\ref{32}) ( or equivalently the analysis of the loop "integrals" ${\cal O}_i$ ) in the fuzzy finite setting as well as in the large $N=L+1$ continuum limit of ordinary  ${\bf S}^2\times{\bf S}^2$ is very complicated. The main difficulty is that we are always ( at every step while we take this particular limit ) dealing with highly non-trivial sums. Furthermore the last term in (\ref{32}) is not manifestly gauge covariant and as a consequence it will not be gauge invariant in the large $N$ limit unless it  vanishes.  The non-covariance of the terms which depend on $F_{ab}^{(12)}$ is on the other hand only due to our choice of background gauge field given in (\ref{16}). Thus gauge covariance can be easily restored in these terms by considering general gauge configurations with non-zero $A_a^{(2)}$.  

The situation is much simpler in the case of one single fuzzy sphere where a  delicate cancellation between  ${\cal O}_4$ and ${\cal O}_3{\Delta}_1$ existed and hence we were able to maintain gauge covariance already in the fuzzy setting.

As it turns out we can show in a straightforward way the existence of a canellation between ${\cal O}_4$ and ${\cal O}_3{\Delta}_1$ on fuzzy  ${\bf S}^2\times{\bf S}^2$ if we consider a different large $N$ limit of the field theory. As opposed to the large $N$ "continuum limit" of commutative ${\bf S}^2\times{\bf S}^2$ we consider instead the large $N$ "noncommutative planar limit" of ${\bf R}^2 \times {\bf R}^2_{\theta}$  with strong noncommutativity $\theta$. In the first stage of this limit sums over the second fuzzy sphere can be converted into integrals over the  noncommutative plane which are easier for analysis in many cases. Strong noncommutativity is crucial since it allows us to freez out all degrees of freedom on the second fuzzy sphere except the zero mode. At the end we will take the continuum limit of the first sphere then the usual flattening limit to obtain ordinary ${\bf R}^2$. 

Thus in taking this planar limit we will treat the two spheres differently. Sums over $k_2$ and $l_2$ ( the second sphere ) will be converted into integrals using the planar limit and then calculated whereas sums over $k_1$ and $l_1$ ( the first sphere ) will be computed first in closed forms ( because it is possible to do that in most cases ) then we take the continuum and  flattening  limits.

From the expressions (\ref{32}),(\ref{6j}) and (\ref{6j+}) we can see that the dependence of ${\cal O}_4(p_1,p_2)-{\cal O}_3(p_1,p_2)p_1(p_1+1)$ on the second sphere is given by the double sum
\begin{eqnarray}
I(k_1l_1;p_1p_2)&=&\sum_{k_2,l_2}\frac{(2k_2+1)}{k_1(k_1+1)+k_2(k_2+1)}\frac{(2l_2+1)}{l_1(l_1+1)+l_2(l_2+1)}[1-(-1)^{R+p_1+p_2}]\left\{\begin{array}{ccc}
                   p_2 & k_2 & l_2 \\
            \frac{L}{2} & \frac{L}{2}  & \frac{L}{2}
                 \end{array}\right\}^2\nonumber\\
&\times &x(k_1k_2l_1l_2;p_1p_2)
\end{eqnarray}
where
\begin{eqnarray}
x(k_1k_2l_1l_2;p_1p_2)
&=&-\frac{[l_1(l_1+1)+l_2(l_2+1)][l_1(l_1+1)-k_1(k_1+1)-p_1(p_1+1)]}{p_1(p_1+1)}\nonumber\\
&-&\frac{[k_2(k_2+1)-l_2(l_2+1)][l_1(l_1+1)-k_1(k_1+1)]}{2p_1(p_1+1)}.\label{367}
\end{eqnarray}
Indeed the difference ${\cal O}_4(p_1,p_2)-{\cal O}_3(p_1,p_2)p_1(p_1+1)$ reads explicitly
\begin{eqnarray}
{\cal O}_4(p_1,p_2)-{\cal O}_3(p_1,p_2)p_1(p_1+1)=4(L+1)^2\sum_{k_1,l_1}(2k_1+1)(2l_1+1)\left\{\begin{array}{ccc}
                   p_1 & k_1 & l_1 \\
            \frac{L}{2} & \frac{L}{2}  & \frac{L}{2}
                 \end{array}\right\}^2I(k_1l_1;p_1p_2)\nonumber\\
\end{eqnarray}
In the planar limit we take $N{\longrightarrow}\infty $ and $  R{\longrightarrow}\infty $ ( where $R$ is the radius of the  spheres ) such that ${\theta}^{'}={\theta}/L =R^2/LN$ ( the noncommutativity parameter ) is kept fixed. Then we will  take the limit ${\theta}^{'}\longrightarrow \infty$. Hence since $N$ is very large we can replace $I$ by the expression
\begin{eqnarray}
I(k_1l_1;p_1p_2)&=&\sum_{k_2,l_2}\frac{(2k_2+1)}{k_1(k_1+1)+k_2(k_2+1)}\frac{(2l_2+1)}{l_1(l_1+1)+l_2(l_2+1)}[1-(-1)^{k_1+l_1+p_1}]\frac{(C_{k_20l_2}^{p_20})^2}{2p_2+1}\nonumber\\
&\times &\frac{x(k_1k_2l_1l_2;p_1p_2)}{L+1}\label{I}
\end{eqnarray}
where we have used the asymptotic behaviour of the $6j$ symbol for very large angular momentum $\frac{N-1}{2}=\frac{L}{2}$ given by \cite{VKM}\footnote{Note that compared with \cite{VKM} the leading behaviour of the $6j$ symbol is taken here to be proportional to $1/(L+1)$ instead of $1/L$ for convenience. This difference is clearly unimportant in the large $N$ limit. }
\begin{eqnarray}
\left\{\begin{array}{ccc}
                   p_2 & k_2 & l_2 \\
            \frac{L}{2} & \frac{L}{2}  & \frac{L}{2}                \end{array}\right\}^2{=}\frac{(C_{k_20l_20}^{p_20})^2}{(L+1)(2p_2+1)}+... \label{soussa1}
\end{eqnarray}
Since $R+p_1+p_2$ must be an odd number ( coming from $[1-(-1)^{R+p_1+p_2}]$ ) we can conclude that $p_1+k_1+l_1$ is also odd because $p_2+k_2+l_2$ must be even  from the properties of the Clebsch-Gordan $C_{k_20l_20}^{p_20}$ .

Furthermore in this large planar limit we will identify any angular momentum $j_2$ on the second fuzzy sphere with the corresponding linear momentum $P_{j_2}$ on the noncommutative plane by the relation $j_2(j_2+1)=R^2P_{j_2}^2=N\theta P_{j_2}^2$. As a consequence all angular momenta $p_2$, $k_2$ and $l_2$ on the second fuzzy sphere can be assumed in this planar limit to be very large compared to $1$. Quantum numbers on the first fuzzy sphere are defined by  a similar formula $j_1(j_1+1)=R^2P_{j_1}^2$. 

Since we will take the planar limit of the second fuzzy sphere in such a way that we will have a strong noncommutativity parameter while we will take the continuum limit ( then the ordinary flattening limit ) of the first fuzzy sphere, we need to manipulate momenta on the two spheres differently. In the first stage we will fix the first fuzzy sphere ( in other words we will fix the planar momenta $P_{j_1}$ ) so the effect of the limit on this sphere can be undone at the end while on the other hand because $0{\leq}P_{j_2}^2{\leq}1/{\theta}^{'}$ we can see that planar momenta $P_{j_2}$ on the second fuzzy sphere approach $0$ as $1/\sqrt{{\theta}^{'}}$ which will simplify our integrals considerably. In the second stage we will take the  continuum limit of the first fuzzy sphere then the ordinary flattening limit so we end up with ${\bf R}^2\times {\bf R}_{\theta}^2$. We will also comment in the next section on the noncommutative planar limit of the first fuzzy sphere in which we end up instead with the space
 ${\bf R}_{\theta}^2\times {\bf R}_{\theta}^2$.


All angular momenta $p_2$, $k_2$ and $l_2$ on the second fuzzy sphere are very large compared to $1$ and thus we can approximate the square of the Clebsch-Gordan $C_{k_20l_20}^{p_20}$\footnote{This is the probability to couple the angluar momenta $k_2$ and $l_2$ with projections equal $0$ to give the angular momentum $p_2$ with projection equal $0$. } by \cite{VKM}
\begin{eqnarray}
\frac{(C_{k_20l_20}^{p_20})^2}{2p_2+1}&{\simeq}&\frac{1}{\pi}\frac{1}{\sqrt{-k_2^4-l_2^4-p_2^4+2k_2^2l_2^2+2k_2^2p_2^2+2l_2^2p_2^2}}.\label{soussa2}
\end{eqnarray}
Let us also remark that from the properties of the Clebsch-Gordan coefficients   we know that $l_2$ must be in the range $k_2-p_2{\leq}l_2{\leq}k_2+p_2$. Hence the sum over $l_2$ in $I$ with $x$ equal to the first term in (\ref{367}) will be given by the integral

\begin{eqnarray}
\sum_{k_2-p_2}^{k_2+p_2}\frac{2l_2+1}{\sqrt{-k_2^4-l_2^4-p_2^4+2k_2^2l_2^2+2k_2^2p_2^2+2l_2^2p_2^2}}&=&
\int_{k_2-p_2}^{k_2+p_2}\frac{d[l_2(l_2+1)]}{\sqrt{[l_2^2-(k_2+p_2)^2][(k_2-p_2)^2-l_2^2]}}\nonumber\\
&=&\int_{(P_{k_2}-P_{p_2})^2}^{(P_{k_2}+P_{p_2})^2}
\frac{dP_{l_2}^2}{\sqrt{[P_{l_2}^2-(P_{k_2}+P_{p_2})^2][(P_{k_2}-P_{p_2})^2-P_{l_2}^2]}}\nonumber\\
&=&\int_{0}^{4P_{k_2}P_{p_2}}\frac{dx}{\sqrt{x(4P_{k_2}P_{l_2}-x)}}\nonumber\\
&=&\pi.\label{pi}
\end{eqnarray}
In above we have used $j=\sqrt{\theta N}P_{j}+...$ with corretions which go to $0$ with $N{\longrightarrow}\infty$. This result is independent of $k_2$ and hence the extra sum over $k_2$ in $I$ will lead to
\begin{eqnarray}
\pi \sum_{k_2=0}^L\frac{2k_2+1}{k_2(k_2+1)+k_1(k_1+1)}=\pi\int_{0}^{\frac{1}{{\theta}^{'}}}\frac{dP_{k_2}^2}{P_{k_2}^2+P_{k_1}^2}=\pi\int_{0}^{1}\frac{dy}{y+{\theta}^{'}P_{k_1}^2}.
\end{eqnarray}
Let us recall that $0{\leq}{\theta}^{'}P_{k_2}^2{\leq}1$ from which we see that the range of $P_{k_2}$ shrinks to $0$ and hence the integral is dominated in this limit by the value $P_{k_2}^2=0$. The above sum ( which is proportional to the contribution  to $I$ coming from setting  $x$ equal to the first term in (\ref{367}) ) is equal 

\begin{eqnarray}
\pi\sum_{k_2=0}^L\frac{2k_2+1}{k_2(k_2+1)+k_1(k_1+1)}=\frac{\pi}{{\theta}^{'}P_{k_1}^2}+...\label{ß}
\end{eqnarray}
In this equation we have also used the limit ${\theta}^{'}{\longrightarrow}\infty $ keeping $P_{k_1}$ fixed. Finally we need  to multiply this result by the factor $-\big(k_1(k_1+1)-l_1(l_1+1)-p_1(p_1+1)\big)(1-(-1)^{k_1+l_1+p_1})/(\pi N p_1(p_1+1))$ in accordance with equation  (\ref{I}) to get the full contribution to $I$ coming from setting  $x$ equal to the first term in (\ref{367}) .

 In this strong noncommutativity planar limit of the second fuzzy sphere the UV and IR  regimes on the corresponding noncommutative plane are one and the same if we choose  not to use dimensionless variables. The UV regime should thus be defined by the momenta which are such that ${\theta}^{'}P_{k_2}^2{\longrightarrow}1$ whereas the IR regime should be defined by the momenta for which ${\theta}^{'}P_{k_2}^2{\longrightarrow}0$ otherwise there will be no distinction between these two regions. Everything is measured here in terms of the noncommutativity parameter ${\theta}^{'}$.

The sum over $l_2$ in $I$ with $x$ equal to the second term in (\ref{367}) will lead on the other hand to a vanishingly small contribution in the limit. Indeed this sum is given by the integral

\begin{eqnarray}
\frac{1}{2}\int_{k_2-p_2}^{k_2+p_2}\frac{d[l_2(l_2+1)]}{\sqrt{[l_2^2-(k_2+p_2)^2][(k_2-p_2)^2-l_2^2]}}\frac{k_2(k_2+1)-l_2(l_2+1)}{l_1(l_1+1)+l_2(l_2+1)}&=&\nonumber\\
-\frac{\pi}{2}+\frac{P_{l_1}^2+P_{k_2}^2}{2}\int_{-1}^{+1}\frac{dx}{\sqrt{1-x^2}}\frac{1}{P_{l_1}^2+P_{k_2}^2+P_{p_2}^2-2P_{k_2}P_{p_2}x}.\label{crso}
\label{443}
\end{eqnarray}
The sum over $k_2$ is given by the integral
\begin{eqnarray}
\sum_{k_2=0}^L\frac{2k_2+1}{k_2(k_2+1)+k_1(k_1+1)}f(P^2_{k_2})=\int_{0}^{\frac{1}{{\theta}^{'}}}\frac{dP_{k_2}^2}{P_{k_2}^2+P_{k_1}^2}f(P^2_{k_2})=\int_{0}^{1}\frac{dy}{y+{\theta}^{'}P_{k_1}^2}f(\frac{y}{{\theta}^{'}}).\label{int}
\end{eqnarray}
The function $f$ from (\ref{crso}) is 
\begin{eqnarray}
f(\frac{y}{{\theta}^{'}})&=&
-\frac{\pi}{2}+\frac{{\theta}^{'}P_{l_1}^2+y}{2}\int_{-1}^{+1}\frac{dx}{\sqrt{1-x^2}}\frac{1}{{\theta}^{'}P_{l_1}^2+y+{\theta}^{'}P_{p_2}^2-2\sqrt{y{\theta}^{'}}P_{p_2}x}\nonumber\\
&=&-\frac{\pi}{2}\frac{P_{p_2}^2}{P_{l_1}^2+P_{p_2}^2}\nonumber\\
&=&-\frac{\pi}{2}\frac{P_{p_2}^2}{P_{l_1}^2}+...\label{pi1}
\end{eqnarray}
In above we have used again the limit ${\theta}^{'}{\longrightarrow}\infty$ then we used the fact that the momentum $P_{k_1}$ on the first fuzzy sphere is fixed wehereas the momentum $P_{p_2}$ on the second fuzzy sphere is such that $0{\leq}P_{p_2}^2{\leq}1/{{\theta}^{'}}$ and thus it goes to zero as $1/\sqrt{{\theta}^{'}}$. We obatin then
\begin{eqnarray}
{\rm equation}~ (\ref{443})=-\frac{\pi}{2}\frac{1}{{\theta}^{'}P_{k_1}^2}\frac{P_{p_2}^2}{P_{l_1}^2}.
\end{eqnarray}
 We can  check that this yields zero contribution to $I$ because of the factor $l_1(l_1+1)-k_1(k_1+1)$ in the second line of (\ref{367}).
This is expected since for $l_2=k_2$ ( which is the value which dominates the sum over $l_2$ ) the second term in (\ref{367}) is zero. This is also expected from the fact that if we set $p_2=0$ then we must have $l_2=k_2$ and hence the second term in (\ref{367}) vanishes.

Therefore the sum $I$ will be dominated in this limit by  the part with $x$ equal to the first term in (\ref{367}) which is given by equation (\ref{ß}) $\times$ $-\big(k_1(k_1+1)-l_1(l_1+1)-p_1(p_1+1)\big)(1-(-1)^{k_1+l_1+p_1})/(N \pi p_1(p_1+1))$ in accordance with equation  (\ref{I}). This yields the difference

\begin{eqnarray}
{\cal O}_4(p_1,p_2)-{\cal O}_3(p_1,p_2)p_1(p_1+1)&=&-4L(L+1)^2\sum_{k_1,l_1}\frac{(2k_1+1)(2l_1+1)}{k_1(k_1+1)}[1-(-1)^{k_1+l_1+p_1}]\nonumber\\
&\times &\left\{\begin{array}{ccc}
                   p_1 & k_1 & l_1 \\
            \frac{L}{2} & \frac{L}{2}  & \frac{L}{2}
                 \end{array}\right\}^2
\frac{l_1(l_1+1)-k_1(k_1+1)-p_1(p_1+1)}{p_1(p_1+1)} \nonumber\\
\end{eqnarray}
By using the result of \cite{ref} we can do the remaining sums over $k_1$ and $l_1$ to obtain

\begin{eqnarray}
{\cal O}_4(p_1,p_2)-{\cal O}_3(p_1,p_2)p_1(p_1+1)&=&8L(L+1).
\end{eqnarray}
 Similarly we can compute the double sum (\ref{c2}) in this planar limit by first converting the sum over $k_2$  into an integral and then performing the sum over $k_1$. The result is as follows 
\begin{eqnarray}
-4\hat{c}_2&=&-8{L}(L+1).
\end{eqnarray}
Therefore we see that in this limit 
\begin{eqnarray}
{\cal O}_4(p_1,p_2)-{\cal O}_3(p_1,p_2)p_1(p_1+1)-4\hat{c}_2=0.
\end{eqnarray}
The effective action in this noncommutative planar limit on noncommutative ${\bf S}^2_N\times{\bf R}^2_{\theta}$  ( the first sphere is still fuzzy ) becomes manifestly gauge-covariant given by
\begin{eqnarray}
\Gamma_2&=&S_2^{(1)}+S_2^{(1,2)}+Tr_LF_{ab}^{(1)}\bigg[\frac{1}{2}\frac{{\cal O}_4}{{\Delta}_1}-\frac{4L(L+1)}{{\Delta}_1}-{\cal O}_F\bigg]F_{ab}^{(1)}-{\epsilon}_{abc}Tr_LF_{ab}^{(1)}\bigg[\frac{1}{2}\frac{{\cal O}_4}{{\Delta}_1}-\frac{4L(L+1)}{{\Delta}_1}\bigg]A_a^{(1)}\nonumber\\
&-&2Tr_LF_{ab}^{(12)}{\cal O}_F F_{ab}^{(12)}.
\end{eqnarray}
For completeness let us also discuss what happens to ${\cal O}_F$ in this noncommutative planar limit of the second fuzzy sphere. The relevant integral  over $k_2$ and $l_2$ is given in this case by
\begin{eqnarray}
I_F(k_1l_1;p_1p_2)
&=&\frac{1}{2\pi N^2 \theta}[1-(-1)^{k_1+l_1+p_1}]\int_{0}^{\frac{1}{{\theta}^{'}}}\frac{dP_{k_2}^2}{P_{k_2}^2+P_{k_1}^2}\int_{-1}^{+1}\frac{dx}{\sqrt{1-x^2}} \frac{1}{P_{l_1}^2+P_{k_2}^2+P_{p_2}^2-2P_{k_2}P_{P_2}x}\nonumber\\
&=&\frac{L}{2 N^2 {\theta}^2}[1-(-1)^{k_1+l_1+p_1}]\frac{1}{(P_{k_1}^2)(P_{l_1}^2+P_{p_2}^2)}+...\nonumber\\
&=&\frac{L}{2}[1-(-1)^{k_1+l_1+p_1}]\frac{1}{k_1(k_1+1)l_1(l_1+1)}+...
\end{eqnarray}
In above we have again neglected corrections proportional to the external momentum $P_{p_2}$ since it goes to $0$ in this limit  as $1/\sqrt{{\theta}^{'}}$.   The remaining sums over $k_1$ and $l_1$ are done in \cite{ref} where we found the result ${\cal O}_F{\longrightarrow}0$ in the continuum limit of the ordinary sphere. Hence the effective action becomes on noncommutative ${\bf S}^2\times {\bf R}^2_{\theta}$   given by
\begin{eqnarray}
\Gamma_2&=&S_2^{(1)}+S_2^{(1,2)}+Tr_LF_{ab}^{(1)}\bigg[\frac{1}{2}\frac{{\cal O}_4}{{\Delta}_1}-\frac{4L(L+1)}{{\Delta}_1}\bigg]F_{ab}^{(1)}-{\epsilon}_{abc}Tr_LF_{ab}^{(1)}\bigg[\frac{1}{2}\frac{{\cal O}_4}{{\Delta}_1}-\frac{4L(L+1)}{{\Delta}_1}\bigg]A_a^{(1)}.\label{pla}\nonumber\\
\end{eqnarray}
The two quantum contributions ( the second and third terms ) are non-zero gauge invariant corrections to the classical action which shows that the quantum theory of $U(1)$ fields on noncommutative ${\bf S}^2\times {\bf R}^2_{\theta}$ is a non-trivial theory as opposed to the  quantum theory of $U(1)$ fields on commutative ${\bf S}^2\times {\bf R}^2$ which is trivial. This indicates the presence of a UV-IR mixing phenomena in this model.

Finally we need to compute ${\cal O}_4$ by converting the sums over $k_2$ and $l_2$ into integrals using the large ${\theta}^{'}$ limit, then performing the remaining sums over $k_1$ and $l_1$. In this case we have
\begin{eqnarray}
I_4(k_1l_1;p_1p_2)
&=&\frac{1}{\pi N }[1-(-1)^{k_1+l_1+p_1}]\int_{0}^{\frac{1}{{\theta}^{'}}}dP_{k_2}^2\int_{-1}^{+1}\frac{dx}{\sqrt{1-x^2}} \frac{1}{P_{l_1}^2+P_{k_2}^2+P_{p_2}^2-2P_{k_2}P_{P_2}x}\nonumber\\
&=&[1-(-1)^{k_1+l_1+p_1}]\frac{L}{l_1(l_1+1)}+...
\end{eqnarray}
Again we have neglected subleading corrections which are proportional to $P_{p_2}$. Finally performing the sums over $k_1$ and $l_1$ yields the answer
\begin{eqnarray}
\frac{1}{2}\frac{{\cal O}_4}{{\Delta}_1}(p_1,p_2)-\frac{4L(L+1)}{{\Delta}_1}&=&\frac{4L(L+1)}{p_1(p_1+1)}\sum_{k=2}^{p_1}\frac{1}{k}.\label{rrrr}
\end{eqnarray}
The flattening limit of the above action is straightforward and we can immediately conclude that the $U(1)$ theory on noncommutative $R^2\times R^2_{\theta}$ is non-trivial as opposed to $U(1)$ theory on $R^2\times R^2$.

\section{The beta function and the planar limit ${\bf R}^2_{\theta}\times {\bf R}^2_{\theta}$}

Let us explain the point about the UV-IR mixing further by taking the planar limit of the first fuzzy sphere in computing the operator ${\cal O}_4$ so we end up  with ${\bf R}^2_{\theta}\times {\bf R}^2_{\theta}$ instead. Modulo the terms involving ${\cal O}_F$ ( which need to be recalculated  ) the above action (\ref{pla}) is still valid in this different limit.  Now from the other expression of ${\cal O}_4$ found in the appendix we have

\begin{eqnarray}
{\cal O}_4(p_1,p_2)&=&4\sum_{k_1,k_2}\frac{(2k_1+1)(2k_2+1)}{k_1(k_1+1)+k_2(k_2+1)}\bigg[1-(-1)^{k_1+k_2+p_1+p_2}(L+1)^2 \times \nonumber\\ 
&&\left\{\begin{array}{ccc}
                   p_1 & \frac{L}{2} & \frac{L}{2} \\
            k_1 & \frac{L}{2}  & \frac{L}{2}
                 \end{array}\right\}\left\{\begin{array}{ccc}
                   p_2 & \frac{L}{2} & \frac{L}{2}\\
            k_2 & \frac{L}{2}  & \frac{L}{2}
                 \end{array}\right\}\bigg].
\end{eqnarray}
The first term inside the bracket results from performing the sum over $l_1$ and $l_2$  in (\ref{6j}) with $[1-(-1)^{R+p_1+p_2}]$ replaced with $1$; this is the planar contribution to the diagram. The second term in the above formula results on the other hand from performing the sum with $[1-(-1)^{R+p_1+p_2}]$ replaced with $-(-1)^{R+p_1+p_2}$; this is the nonplanar contribution. In the commutative these two contributions are equal and hence they cancel each other.

As it turns out 
we can use in the large $L$ limit ( for $p_1,p_2<<\frac{L}{2}$ and $0{\leq}k_1,k_2{\leq}L$ ) the same approximation used in \cite{uvir}, namely
\begin{eqnarray}
\left\{ \begin{array}{ccc}
        p_1 & \frac{L}{2} & \frac{L}{2} \\
    k_1 & \frac{L}{2} & \frac{L}{2}
    \end{array} \right\}{\simeq}\frac{(-1)^{L+p_{1}+k_1}}{L+1}P_{p_1}(1-\frac{2k_1^2}{L^2})~,~\left\{ \begin{array}{ccc}
        p_2 & \frac{L}{2} & \frac{L}{2} \\
    k_2 & \frac{L}{2} & \frac{L}{2}
    \end{array} \right\}{\simeq}\frac{(-1)^{L+p_{2}+k_2}}{L+1}P_{p_2}(1-\frac{2k_2^2}{L^2})
\end{eqnarray}
To obtain
\begin{eqnarray}
{\cal O}_4(p_1,p_2)&=&4\int_{0}^{\frac{1}{{\theta}^{'}}}\frac{dP_{k_1}^2dP_{k_2}^2}{P_{k_1}^2+P_{k_2}^2}\bigg[1-P_{p_1}\big(1-\frac{2\theta}{L}P_{k_1}^2\big)P_{p_2}\big(1-\frac{2\theta}{L}P_{k_2}^2\big)\bigg].
\end{eqnarray}
$P_{p}$ are the Legendre polynomials.
For $p_1>>1$ and ${\theta}^{'}P_{k_1}^2<<1$ we can also use the approximations used in \cite{uvir}, viz
\begin{eqnarray}
P_{p_1}\big(1-\frac{2\theta}{L}P_{k_1}^2\big)&=&J_0(2\theta P_{p_1}P_{k_1})+...=\int \frac{d{\phi}_1}{2\pi}e^{2i\theta P_{p_1}P_{k_1}\cos{\phi}_1}.
\end{eqnarray}
By rotational invariance we have ${\theta}P_{p_1}^{\mu}B_{\mu \nu}P_{k_1}^{\nu}={\theta}P_{p_1}(P_{k_1}cos{\phi}_1)$ ( with $B_{12}=-1$ ) where we have chosen the $2-$dimensional external momentum $P_{p_1}$ to lie in the $y-$direction and  ${\phi}_1$ is the angle between the internal momentum $\vec{P_{k_1}}$ and the $x-$axis. Thus we obatin ( with $d^2P_{k_1}=P_{k_1}dP_{k_1}d{\phi}_1$ and $d^2P_{k_2}=P_{k_2}dP_{k_2}d{\phi}_2$)
\begin{eqnarray}
{\cal O}_4(p_1,p_2)&=&\frac{4}{{\pi}^2}\int_{0}^{\frac{L}{{\theta}}}\frac{d^2{P}_{k_1}d^2{P}_{k_2}}{P_{k_1}^2+P_{k_2}^2}\bigg[1-e^{2i\theta[P_{p_1}BP_{k_1}+P_{p_2}BP_{k_2}]}\bigg].\label{770}
\end{eqnarray}
The second term is precisley the canonical non-planar 2-point function on noncommutative ${\bf R}^4_{\theta}$ with Euclidean metric ${\bf R}^2_{\theta}\times {\bf R}^2_{\theta}$ whereas the first term is the planar contribution. Two important remarks are now in order. $1)$ The external legs of this 2-point function are two curvature tensors $F_{ab}^{(1)}$ for the third term of (\ref{pla}) and one curvature tensor $F_{ab}^{(1)}$ and one gauge field ${\epsilon}_{abc}A_c^{(1)}$ for the fourth term of (\ref{pla}). $2)$ These contributions are not necessarily identical to what we usually obtain on noncommutative ${\bf R}_{\theta}^4$. This is not surprising since we obtained this noncommutative ${\bf R}_{\theta}^4$ in a very special way by scaling fuzzy ${\bf S}^2\times {\bf S}^2$. The scaled commutation relations can be checked to be those of noncommutative ${\bf R}^4_{\theta}$. The structure of the phases indicates also that we are indeed dealing with a noncommutative ${\bf R}^4_{\theta}$. Comparing (\ref{770}) with the first equation of section $4.${\rm D} of \cite{others1} we can see that (\ref{770}) corresponds to the second term of that equation which is proportional to the metric ${\eta}_{ij}$.

The above integral shows a divergence at zero momentum. This is only an artifact of the approximation used  above. The regularized value of ${\cal O}_4$ is given by (\ref{rrrr}). If we take  the planar limit of the first fuzzy sphere by rewriting this result as
\begin{eqnarray}
\frac{1}{2}\frac{{\cal O}_4}{{\Delta}_1}(p_1,p_2)-\frac{4L(L+1)}{{\Delta}_1}
&=&\frac{4}{{\theta}^{'}P_{p_1}^2}\ln \frac{R|P_{p_1}|}{2}.
\end{eqnarray}
Then We recover the usual logarithmic divergence when  $R{\longrightarrow}0$ ( i.e ${\theta}^{'}{\longrightarrow}0$ ) or equivalently $P_{p_1}{\longrightarrow}0$. 

The renormalized $U(1)$ gauge coupling constant $g^2(P)$ ( which is obviously momentum dependent ) can be easily read from the above discussion. We obatin immediately the formula
\begin{eqnarray}
\frac{1}{4g^2(P)}=\frac{1}{4g^2}+\frac{2}{{\theta}^{'}P^2}\ln \frac{R^2P^2}{4}.
\end{eqnarray}
$P$ is the momentum on the first NC plane ${\bf R}^2_{\theta}$ which is in the range $0{\leq}{\theta}^{'}P^2{\leq}1$. Looking at high momenta ${\mu}^2$ in the UV region near the cut-off $1/{\theta}^{'}$ we can see that $\frac{2}{{\theta}^{'}P^2}\ln \frac{R^2P^2}{4}$ is approximated by $2\ln \frac{R^2P^2}{4}$ and hence in this regime the above formula reduces to
\begin{eqnarray}
\frac{1}{4g^2(\mu)}=\frac{1}{4g^2}+2\ln \frac{R^2{\mu}^2}{4}
\end{eqnarray}
or equivalently
\begin{eqnarray}
\mu \frac{\partial g(\mu)}{\partial \mu}=-8g^3(\mu)
\end{eqnarray}
This is ( upto a multiplicative factor ) the same beta function derived for noncommutative $U(1)$ gauge theory in \cite{martin,badis1}.  The most important
things are the negative sign and the cube power which come naturally out of the model.
Thus the strong noncommutativity limit considered in this note captures already most of the essential feature of noncommutative $U(1)$ gauge theory on ${\bf R}^4_{\theta}$. This also shows explicitly how  large $N$ fuzzy ${\bf S}^2\times {\bf S}^2$ acts as a regulator of noncommutative ${\bf R}^4_{\theta}$. 

Restoring $SO(4)$ covariance is straightforward. We only need to consider a background gauge configuration with both $A_a^{(1)}$ and $A_a^{(2)}$ non-zero. There will be extra terms to be computed in the effective action but fortunately all of them ( even those which mixe $A_a^{(1)}$ and $A_a^{(2)}$ ) are of the same strucutre as those already considered in this article. See the effective action (\ref{dlta}) . Their analysis will be therefore easy and there will no additional physics to be learned from this calculation.

The last point is with regard to the mass terms in the original model (\ref{model}). In two dimensions the presence of such terms with $m{\longrightarrow}\infty $ causes the UV-IR mixing to disappear. Loosely speaking this can be traced to the dimensionality of the space. As we have already explained the effect of these terms in the large mass limit is only to project out the scalar normal components from the theory and hence effectively reduce the three dimensional trace in $\frac{1}{2}Tr_3TR\log {\Omega}$ to a two dimensional trace. Taking also the ghost contribution $-TR\log {\cal D}^2$ into account we see that the effective action will only consist in terms depending on the curvature which as we have shown in \cite{ref} go to zero in the limit anyway. This scenario does not happen here in $4$ dimensions for the obvious reason that we have in this case the gauge contribution $\frac{1}{2}Tr_6TR\log {\Omega}$  while the ghost contribution does not change and it is still given formally by $-TR\log {\cal D}^2$ . See equation (\ref{effaction}). Thus with the presence of the mass terms ( even if we let the mass goes to infinity ) we expect the UV-IR mixing to persist in $4$ dimensions as opposed to $2$ dimensions. 
\section{Effective potential on ${\bf S}^2\times {\bf S}^2$ and topology change}

Let us redo the calculation of the effective potential on  fuzzy ${\bf S}^2\times {\bf S}^2$ done originally in the first reference of \cite{S2S2}. In here we will also consider  the case when the two  spheres have different radii and hence the configurations of interest are given by
\begin{eqnarray}
D_a^{(1)}={\phi}_1L_a^{(1)}~,~D_a^{(2)}={\phi}_2L_a^{(2)}.
\end{eqnarray}
The starting point is the effective action (\ref{effaction}) with $\xi=1$ and $m$ small. After some calculation we can show that the value of the effective action in these configurations is given by

\begin{eqnarray}
V_{\rm eff}({\phi}_1,{\phi}_2)={\Gamma}\big({\phi}_1L_a^{(1)},{\phi}_2L_a^{(2)}\big)&=&\frac{L(L+2)}{2g^2}\bigg[\frac{1}{4}{\phi}_1^4-\frac{1}{3}{\phi}_1^3+\frac{m^2}{4}({\phi}_1^2-1)^2]\nonumber\\
&+&\frac{L(L+2)}{2g^2}\bigg[\frac{1}{4}{\phi}_2^4-\frac{1}{3}{\phi}_2^3+\frac{m^2}{4}({\phi}_2^2-1)^2]\nonumber\\
&+&2{\rm TR}\log\bigg[{\phi}_1^2({\cal L}_a^{(1)})^2+{\phi}_2^2({\cal L}_a^{(2)})^2\bigg]+....
\end{eqnarray}
The only interaction between the two spheres is in the quantum contribution.

There are several possibilities to be considered here. First of all if we insist on the full $SO(4)$ rotational invariance then we must set ${\phi}_1={\phi}_2\equiv \phi$. In this case we can compute that the above potential will admit a stable minimum ( in other words a solution $\phi$ to the equation of motion wille exist ) for all values of $g^2$ which are less than the critical value ( see the first reference of \cite{S2S2} )
\begin{eqnarray}
{g_*^2L^2}|_1=\frac{m^2+\sqrt{2}-1}{16}.
\end{eqnarray}
Below this value we have $\phi {\simeq}1$ ( "the fuzzy ${\bf S}^2\times {\bf S}^2$ phase ") whereas above this value we have $\phi {\longrightarrow}0$  ( "the matrix phase" ). In the fuzzy ${\bf S}^2\times {\bf S}^2$  phase the field theory is a $U(1)$ gauge theory at least in the very weak coupling region. We suspect that the gauge group structure inside the fuzzy ${\bf S}^2\times {\bf S}^2$ phase will change at some point ( which means another phase transition ) from $U(1)$ to $U((L+1)^2)$ in analogy to what happened on a single fuzzy sphere where the gauge group changed from $U(1)$ to $U(L+1)$ inside the fuzzy sphere phase. Indeed the dynamics inside the matrix phase is given by a $U((L+1)^2)$ gauge theory on a point so the expectation that the gauge group will change inside the fuzzy ${\bf S}^2\times {\bf S}^2$ at some coupling before we reach the matrix phase is natural. The most important point in all this physics is the topology change ${\bf S}^2\times {\bf S}^2{\longrightarrow}\{0\}$ which seems to be related to the UV-IR mixing phenomena.

But there is more. If we do not insist on $SO(4)$ rotational invariance then we can consider the configurations with ${\phi}_1=1$ and ${\phi}_2\equiv {\phi}$ ( or the other way around ). Then similarly to above we can compute that the potential will admit a stable minimum for all values of $g^2$ which are less than the critical value 
\begin{eqnarray}
{g_*^2L^2}|_2=\frac{m^2+\sqrt{2}-1}{32}.
\end{eqnarray}
This is half the original value ${g_*^2L^2}|_1$. 
Below this value we have again "the fuzzy ${\bf S}^2\times {\bf S}^2$ phase " where $\phi {\simeq}1$ whereas above this value we have $\phi {\longrightarrow}0$   which is now a "fuzzy ${\bf S}^2$ phase ".  The  topology change here is seen to be ${\bf S}^2\times {\bf S}^2{\longrightarrow}{\bf S}^2$.

Putting the two facts together we have the following picture. For values of $g$ below ${g_*}|_2$ we have a fuzzy ${\bf S}^2\times {\bf S}^2$ while for values of $g$ between ${g_*}|_2$ and ${g_*}|_1$ we have a single fuzzy sphere and for values of $g$ above ${g_*}|_1$ we have a single point. Thus the topology change obtained in this model is ${\bf S}^2\times {\bf S}^2{\longrightarrow}{\bf S}^2{\longrightarrow}\{0\}$. Remark that the critical values become large for large values of the mass $m$ which makes the transitions and as a consequence the topology change harder to reach from small couplings.

\section{Conclusion}

In this article we have calculated   the one-loop quantum correction of $U(1)$ gauge fields on  fuzzy ${\bf S}^2\times {\bf S}^2$. In the large $N$  planar limit we have shown  the existence of a gauge invariant UV-IR mixing. We have also computed the beta function. In the strong noncommutativity limit considered here most of the essensial features of  noncommutative $U(1)$ gauge theory on the Moyal-Weyl  $R^4_{\theta}$ emerged. In this sense we have explicitly shown that  large $N$  fuzzy ${\bf S}^2\times {\bf S}^2$ can be used as  regulator of  gauge theory on $R^4_{\theta}$.

In this model we have also shown from the computation of the effective potential the existence of  ( first order)  phase transitions $1)$ from fuzzy ${\bf S}^2\times{\bf S}^2$ to ${\bf S}^2$ and then from ${\bf S}^2$ to  a single point ( matrix phase ) or $2)$ directly from  fuzzy ${\bf S}^2\times {\bf S}^2$ to a matrix phase. This last transition is also rotationally invariant. We argued that this topology change is related to the perturbative UV-IR mixing. This picture seems to be consistent in $2$ dimensions. The transitions can be removed if we take the mass of the normal scalar fields to infinity however the UV-IR mixing in this case ( as opposed to $2$ dimensions ) persists.

Since one of our main goal is to have a nonperturbative regularization of $U(1)$ gauge theory in $4$ dimensions we must find a way to get rid of ( or at least understand better ) the UV-IR mixing and the matrix phases. The inclusion of fermions in this model is a very important issue since it would give us a nonperturbative approach to QED ( or QCD for higher gauge groups ). We also think that adding fermions will remove to a large extent the topology change observed in this model. Fuzzy perturbation theory involving fermions will be reported hopefully soon \cite{in-progress}.

\paragraph{Acknowledgements}
R. Delgadillo-Blando would like
to thank Denjoe O'Connor for his supervision throughout the course of
this study. The work of R.D.B. is supported by CONACYT M\'exico. Badis Ydri would like also to thank Denjoe O'Connor for discussions and critical comments.

The work of Badis Ydri is supported by a Marie Curie Fellowship from The Commision of the European Communities ( The Research Directorate-General ) under contract number MIF1-CT-2006-021797. B.Y would like also to thank the staff at Humboldt-Universitat zu Berlin for their help and support. In particular he would like to thank Michael Muller-Preussker and Wolfgang Bietenholz.


\appendix
\section{Fuzzy perturbation theory}

\paragraph{The tadpole diagram}

We will use the following identities

\begin{eqnarray}
\yyd_{l_1m_1;l_2m_2}=(-1)^{m_1+m_2}\yy_{l_1-m_1;l_2-m_2}~,~\Trl(\yy_{k_1n_1;k_2n_2}\yyd_{l_1m_1;l_2m_2})=\delta_{k_1l_1}\delta_{k_2l_2}\delta_{n_1m_1}\delta_{n_2m_2}.
\end{eqnarray}
\begin{eqnarray}
[\yy_{k_1n_1;k_2n_2},\yy_{l_1m_1;l_2m_2}]&=&\sum_{l_3m_3}\sum_{l_3^{'}m_3^{'}}
{\Omega}_{k_1n_1k_2n_2;l_1m_1l_2m_2}^{l_3m_3l_3^{'}m_3^{'}}\yy_{l_3m_3;l_3^{'}m_3^{'}}\nonumber\\
&=&(L+1)\sqrt{(2l_1+1)(2l_2+1)(2k_1+1)(2k_2+1)}\nonumber\\
&&\times \sum_{l_3m_3}\sum_{l'_3m_3^{'}}(-1)^{l_3+l'_3}[1-(-1)^{R+l_3+l'_3}]
                                \left\{ \begin{array}{ccc}
                   k_1 & l_1 & l_3 \\
            \frac{L}{2} & \frac{L}{2}  & \frac{L}{2}
                 \end{array} \right\}
            \left\{ \begin{array}{ccc}
       k_2 & l_2 & l'_3 \\
              \frac{L}{2} & \frac{L}{2}  & \frac{L}{2} 
    \end{array} \right\}\nonumber\\
&&\times C^{l_3\;m_3}_{k_1n_1\;l_1m_1}C^{l'_3\;m'_3}_{k_2n_2\;l_2m_2}\yy_{l_3m_3;l'_3m'_3}~\nonumber\\
&&~R=k_1+l_1+k_2+l_2,
\end{eqnarray}
and
\begin{eqnarray}
&&[L^{(1)}_{\mu},\yy_{l_1m_1;l_2m_2}]=\sqrt{l_1(l_1+1)}\;C^{l_1\;m_1+\mu}_{l_1m_1\;1\mu}\;\yy_{l_1m_1+\mu;l_2m_2}.
\end{eqnarray}
A straightforward calculation yields
\begin{eqnarray}
{\gamma}_a(k_1n_1,k_2n_2)={\eta}_a^{\mu}(-1)^{\mu}\sum_{l_1m_1,l_2m_2}\frac{\sqrt{l_1(l_1+1)}}{l_1(l_1+1)+l_2(l_2+1)}C_{l_1-m_1,1\mu}^{l_1-m_1+\mu}{\Omega}_{k_1n_1k_2n_2;l_1m_1l_2m_2}^{l_1m_1-\mu l_2m_2}.
\end{eqnarray}
The coefficients ${\eta}_a^{\mu}$ satisfy ${\eta}_a^{\mu}{\eta}_a^{\nu}=(-1)^{\mu}{\delta}_{\mu+\nu,0},a=1,2,3,\mu=0,+1,-1$. The sums over $m_1$ and $m_2$ can be done using the identities
\begin{eqnarray}
\sum_{m_1}C_{l_1-m_1 1\mu}^{l_1-m_1+\mu}C_{k_1n_1l_1m_1}^{l_1m_1-\mu}=\frac{2l_1+1}{3}{\delta}_{k_11}{\delta}_{n_1,-\mu}~,~\sum_{m_2}C_{k_2n_2 l_2m_2}^{l_2m_2}=(2l_2+1){\delta}_{k_20}{\delta}_{n_20}.
\end{eqnarray}
We find
\begin{eqnarray}
{\gamma}_a(k_1n_1,k_2n_2)=\frac{1}{3}{\eta}_a^{\mu}(-1)^{\mu}{\delta}_{k_1,1}{\delta}_{k_2,0}{\delta}_{n_1,-\mu}{\delta}_{n_2,0}\sum_{l_1,l_2}\frac{(2l_1+1)(2l_2+1)}{l_1(l_1+1)+l_2(l_2+1)}\sqrt{l_1(l_1+1)}{\omega}_{l_1l_2},
\end{eqnarray}
where

\begin{eqnarray}
{\omega}_{l_1l_2}
&=&2(L+1)\sqrt{3(2l_1+1)(2l_2+1)}(-1)^{l_1+l_2}
                                \left\{ \begin{array}{ccc}
                   1 & l_1 & l_1 \\
            \frac{L}{2} & \frac{L}{2}  & \frac{L}{2}
                 \end{array} \right\}
            \left\{ \begin{array}{ccc}
       0 & l_2 & l_2 \\
              \frac{L}{2} & \frac{L}{2}  & \frac{L}{2} 
    \end{array} \right\}\nonumber\\
&=&-2\sqrt{3}\frac{\sqrt{l_1(l_1+1)}}{\sqrt{L(L+2)}},
\end{eqnarray}
where we have used the two $6$j symbols
\begin{eqnarray}
 \left\{\begin{array}{ccc}
                   1 & l_1 & l_1 \\
            \frac{L}{2} & \frac{L}{2}  & \frac{L}{2}
                 \end{array}\right\}=\frac{(-1)^{L+l_1+1}\sqrt{l_1(l_1+1)}}{\sqrt{L(L+1)(L+2)(2l_1+1)}}~,~ \left\{\begin{array}{ccc}
                   0 & l_2 & l_2 \\
            \frac{L}{2} & \frac{L}{2}  & \frac{L}{2}
                 \end{array}\right\}=\frac{(-1)^{L+l_2}}{\sqrt{(L+1)(2l_2+1)}}.
\end{eqnarray}
Hence
\begin{eqnarray}
{\gamma}_a(k_1n_1,k_2n_2)&=&-\frac{2}{\sqrt{3L(L+2)}}{\eta}_a^{\mu}(-1)^{\mu}{\delta}_{k_1,1}{\delta}_{k_2,0}{\delta}_{n_1,-\mu}{\delta}_{n_2,0}\sum_{l_1=1}^L\sum_{l_2=0}^L\frac{(2l_1+1)(2l_2+1)}{l_1(l_1+1)+l_2(l_2+1)}l_1(l_1+1)\nonumber\\
&=&-\frac{(L+1)^4-1}{\sqrt{3L(L+2)}}{\eta}_a^{\mu}(-1)^{\mu}{\delta}_{k_1,1}{\delta}_{k_2,0}{\delta}_{n_1,-\mu}{\delta}_{n_2,0}.
\end{eqnarray}
The Tadpole diagram is therefore given by
\begin{eqnarray}
2TR\frac{1}{\Delta}\bigg(\mathcal{L}_{a}^{(1)}\mathcal{A}_{a}^{(1)}+\mathcal{A}_{a}^{(1)}\mathcal{L}_{a}^{(1)} \bigg)&=&4\frac{(L+1)^4-1}{\sqrt{3L(L+2)}}A_{-\mu}^{(1)}(1-\mu,00)\nonumber\\
&=&8\frac{(L+1)^4-1}{L(L+2)}Tr_LA_a^{(1)}L_a^{(1)}\nonumber\\
&=&8\frac{(L+1)^4-1}{L(L+2)}\sqrt{c_2}Tr_L{\Phi}_1-4\frac{(L+1)^4-1}{L(L+2)}Tr_L(A_a^{(1)})^2.\nonumber\\
\end{eqnarray}

\paragraph{The $4-$vertex correction}

We can immediately compute
\begin{eqnarray}
&&\sum_{m_1,m_2}(-1)^{m_1+m_2}Tr_L[\hat{\mathcal{Y}}_{k_1^{'}n_1^{'};k_2^{'}n_2^{'}},\hat{\mathcal{Y}}_{l_1-m_1;l_2-m_2}][\hat{\mathcal{Y}}_{k_1n_1;k_2n_2},\hat{\mathcal{Y}}_{l_1m_1;l_2m_2}]=\nonumber\\
&&\sum_{l_3,l_3^{'}}\sum_{m_1,m_2}(-1)^{m_1+m_2}\sum_{m_3,m_3^{'}}(-1)^{m_3+m_3^{'}}{\Omega}_{k_1^{'}n_1^{'}k_2^{'}n_2^{'},l_1-m_1l_2-m_2}^{l_3m_3l_3^{'}m_3^{'}}{\Omega}_{k_1n_1k_2n_2,l_1m_1l_2m_2}^{l_3-m_3l_3^{'}-m_3^{'}}.
\end{eqnarray}
The sums over $m_1,m_2,m_3^{'}$ and $m_3^{'}$ can be done using the identity
\begin{eqnarray}
\sum_{m_1m_2}(-1)^{m_1+m_2}C^{l_2m_2}_{p_1n_1l_1-m_1}C^{l_2-m_2}_{p_2n_2l_1m_1}=\frac{2l_2+1}{\sqrt{(2p_1+1)(2p_2+1)}}(-1)^{n_1}(-1)^{l_1+l_2+p_1}{\delta}_{p_1p_2}{\delta}_{n_1,-n_2}\nonumber\\
\end{eqnarray}
We obtain ( with ${\delta}_{kk^{'}}=(-1)^{n_1+n_2}{\delta}_{k_1k_1^{'}}{\delta}_{k_2k_2^{'}}{\delta}_{n_1,-n_1^{'}}{\delta}_{n_2,-n_2^{'}}$ and $R=l_1+l_2+k_1+k_2$)
\begin{eqnarray}
-2(L+1)^2(2l_1+1)(2l_2+1){\delta}_{kk^{'}}&&\sum_{l_3,l_3^{'}}(2l_3+1)(2l_3^{'}+1)(1-(-1)^{R+l_3+l_3^{'}})\left\{\begin{array}{ccc}
                   k_1 & l_1 & l_3 \\
            \frac{L}{2} & \frac{L}{2}  & \frac{L}{2}
                 \end{array}\right\}^2\times\nonumber\\
& &\left\{\begin{array}{ccc}
                   k_2 & l_2 & l_3^{'} \\
            \frac{L}{2} & \frac{L}{2}  & \frac{L}{2}
                 \end{array}\right\}^2
\end{eqnarray}
or equivalently ( by using identities $5.2$ and $5.3$ of \cite{ref} ) 
\begin{eqnarray}
-2(L+1)^2(2l_1+1)(2l_2+1){\delta}_{kk^{'}}\bigg[\frac{1}{(L+1)^2}-(-1)^R\left\{\begin{array}{ccc}
                   k_1 & \frac{L}{2} & \frac{L}{2} \\
            l_1 & \frac{L}{2}  & \frac{L}{2}
                 \end{array}\right\}\left\{\begin{array}{ccc}
                   k_2 & \frac{L}{2} & \frac{L}{2}\\
            l_2 & \frac{L}{2}  & \frac{L}{2}
                 \end{array}\right\}\bigg].
\end{eqnarray}
The $4-$vertex correction is therefore given by
\begin{eqnarray}
\sum_{p_1s_1,p_2s_2}(-1)^{s_1+s_2}A_a^{(1)}(p_1s_1,p_2s_2)A_a^{(1)}(p_1-s_1,p_2-s_2){\cal O}_4(p_1,p_2)
\end{eqnarray}
where
\begin{eqnarray}
{\cal O}_4(p_1,p_2)&=&4\sum_{k_1,k_2}\frac{(2k_1+1)(2k_2+1)}{k_1(k_1+1)+k_2(k_2+1)}\bigg[1-(-1)^{k_1+k_2+p_1+p_2}(L+1)^2\left\{\begin{array}{ccc}
                   p_1 & \frac{L}{2} & \frac{L}{2} \\
            k_1 & \frac{L}{2}  & \frac{L}{2}
                 \end{array}\right\}\left\{\begin{array}{ccc}
                   p_2 & \frac{L}{2} & \frac{L}{2}\\
            k_2 & \frac{L}{2}  & \frac{L}{2}
                 \end{array}\right\}\bigg]\nonumber\\
&=&4(L+1)^2\sum_{k_1,k_2}\sum_{l_1,l_2}\frac{(2k_1+1)(2k_2+1)}{k_1(k_1+1)+k_2(k_2+1)}\frac{(2l_1+1)(2l_2+1)}{l_1(l_1+1)+l_2(l_2+1)}[1-(-1)^{R+p_1+p_2}]\nonumber\\
&\times &\left\{\begin{array}{ccc}
                   p_1 & k_1 & l_1 \\
            \frac{L}{2} & \frac{L}{2}  & \frac{L}{2}
                 \end{array}\right\}^2\left\{\begin{array}{ccc}
                   p_2 & k_2 & l_2 \\
            \frac{L}{2} & \frac{L}{2}  & \frac{L}{2}
                 \end{array}\right\}^2\big(l_1(l_1+1)+l_2(l_2+1)\big).\label{39}
\end{eqnarray}

\paragraph{The $F-$vertex contribution}

The 3-vertex correction corresponding to the curvature $F_{ab}^{(1)}$ is given by 

\begin{eqnarray}
&&TR\frac{1}{{\Delta}}\mathcal{F}^{(1)}_{ab}\frac{1}{{\Delta}}\mathcal{F}^{(1)}_{ab}=\nonumber\\ 
&&\sum_{k}\sum_{l}\frac{Tr_L F^{(1)}_{ab}[\yy_{k_1n_1;k_2n_2},\yyd_{l_1m_1;l_2m_2}]Tr_L F^{(1)}_{ab}[\yy_{l_1m_1;l_2m_2},\yyd_{k_1n_1;k_2n_2}]}{[k_1(k_1+1)+k_2(k_2+1)][l_1(l_1+1)+l_2(l_2+1)]}.
\end{eqnarray}
In above the notation is $k=(k_1n_1,k_2n_2)$, $l=(l_1m_1,l_2m_2)$. We need to compute
\begin{eqnarray}
\sum_{n_1,n_2}\sum_{m_1,m_2}(-1)^{n_1+n_2+m_1+m_2}\frac{Tr_L \yy_{p_1s_1;p_2s_2}[\yy_{k_1n_1;k_2n_2},\yy_{l_1-m_1;l_2-m_2}]Tr_L \yy_{q_1t_1;q_2t_2}[\yy_{l_1m_1;l_2m_2},\yy_{k_1-n_1;k_2-n_2}]}{[k_1(k_1+1)+k_2(k_2+1)][l_1(l_1+1)+l_2(l_2+1)]}&=&\nonumber\\
\sum_{n_1,n_2}\sum_{m_1,m_2}(-1)^{n_1+n_2+m_1+m_2}\frac{(-1)^{s_1+s_2+t_1+t_2}{\Omega}_{k_1n_1k_2n_2;l_1-m_1l_2-m_2}^{p_1-s_1p_2-s_2}{\Omega}_{l_1m_1l_2m_2;k_1-n_1k_2-n_2}^{q_1-t_1q_2-t_2}}{[k_1(k_1+1)+k_2(k_2+1)][l_1(l_1+1)+l_2(l_2+1)]}.\nonumber\\
\end{eqnarray}
By observing that ${\Omega}_{k_1n_1k_2n_2;l_1-m_1l_2-m_2}^{p_1-s_1p_2-s_2}=\tilde{\Omega}_{k_1k_2;l_1l_2}^{p_1p_2}C_{k_1n_1l_1-m_1}^{p_1-s_1}C_{k_2n_2l_2-m_2}^{p_2-s_2}$, etc with an obvious definition for $\tilde{\Omega}$ and then using the identity
\begin{eqnarray}
\sum_{n_1,m_1}C_{k_1n_1l_1-m_1}^{p_1-s_1}C_{l_1m_1k_1-n_1}^{q_1-t_1}(-1)^{n_1+m_1}=(-1)^{t_1}{\delta}_{q_1p_1}{\delta}_{t_1,-s_1}
\end{eqnarray}
we obtain
\begin{eqnarray}
&{\delta}_{qp}&\frac{\tilde{\Omega}_{k_1k_2;l_1l_2}^{p_1p_2}\tilde{\Omega}_{l_1l_2;k_1k_2}^{p_1p_2}}{[k_1(k_1+1)+k_2(k_2+1)][l_1(l_1+1)+l_2(l_2+1)]}=2(L+1)^2{\delta}_{qp}\nonumber\\
&\times &\frac{(2k_1+1)(2k_2+1)(2l_1+1)(2l_2+1)}{[k_1(k_1+1)+k_2(k_2+1)]
[l_1(l_1+1)+l_2(l_2+1)]}[1-(-1)^{R+p_1+p_2}] \left\{ \begin{array}{ccc}
                   k_1 & l_1 & p_1 \\
            \frac{L}{2} & \frac{L}{2}  & \frac{L}{2}
                 \end{array} \right\}^2\left\{ \begin{array}{ccc}
                   k_2 & l_2 & p_2 \\
            \frac{L}{2} & \frac{L}{2}  & \frac{L}{2}
                 \end{array} \right\}^2\nonumber\\
\end{eqnarray}
Hence the $F$-vertex correction is equal to 

\begin{eqnarray}
TR\frac{1}{{\Delta}}\mathcal{F}^{(1)}_{ab}\frac{1}{{\Delta}}\mathcal{F}^{(1)}_{ab}&=&\sum_{p_1s_1p_2s_2}(-1)^{s_1+s_2}F_{ab}^{(1)}(p_1s_1p_2s_2)F_{ab}^{(1)}(p_1-s_1p_2-s_2){\cal O}_F(p_1,p_2),\nonumber\\
\end{eqnarray}
where

\begin{eqnarray}
{\cal O}_F(p_1,p_2)&=&2(L+1)^2\sum_{k_1,k_2}\sum_{l_1,l_2}
\frac{(2k_1+1)(2k_2+1)(2l_1+1)(2l_2+1)}{[k_1(k_1+1)+k_2(k_2+1)]
[l_1(l_1+1)+l_2(l_2+1)]}\nonumber\\
&\times &[1-(-1)^{R+p_1+p_2}] \left\{ \begin{array}{ccc}
                   k_1 & l_1 & p_1 \\
            \frac{L}{2} & \frac{L}{2}  & \frac{L}{2}
                 \end{array} \right\}^2\left\{ \begin{array}{ccc}
                   k_2 & l_2 & p_2 \\
            \frac{L}{2} & \frac{L}{2}  & \frac{L}{2}
                 \end{array} \right\}^2.\label{46}
\end{eqnarray}

\paragraph{The $3-$vertex correction}

This is given by
\begin{eqnarray}
TR\frac{1}{\Delta}\bigg(\mathcal{L}_{a}^{(1)}\mathcal{A}_{a}^{(1)}+\mathcal{A}_{a}^{(1)}\mathcal{L}_{a}^{(1)}\bigg)\frac{1}{\Delta}\bigg(\mathcal{L}_{a}^{(1)}\mathcal{A}_{a}^{(1)}+\mathcal{A}_{a}^{(1)}\mathcal{L}_{a}^{(1)}\bigg)&=&\nonumber\\
2\sum_{k_1n_1,k_2n_2}\sum_{l_1m_1,l_2m_2}(-1)^{\mu+\nu}\frac{Tr_L[L_{\mu}^{(1)},\hat{\mathcal{Y}}_{k_1n_1;k_2n_2}][A_{-\mu}^{(1)},\yyd_{l_1m_1;l_2m_2}]Tr_L[L_{\nu}^{(1)},\hat{\mathcal{Y}}_{l_1m_1;l_2m_2}][A_{-\nu}^{(1)},\yyd_{k_1n_1;k_2n_2}]}{[k_1(k_1+1)+k_2(k_2+1)][l_1(l_1+1)+l_2(l_2+1)]}&+&\nonumber\\
2\sum_{k_1n_1,k_2n_2}\sum_{l_1m_1,l_2m_2}(-1)^{\mu+\nu}\frac{Tr_L[L_{\mu}^{(1)},\hat{\mathcal{Y}}_{k_1n_1;k_2n_2}][A_{-\mu}^{(1)},\yyd_{l_1m_1;l_2m_2}]Tr_L[A_{-\nu}^{(1)},\hat{\mathcal{Y}}_{l_1m_1;l_2m_2}][L_{\nu}^{(1)},\yyd_{k_1n_1;k_2n_2}]}{[k_1(k_1+1)+k_2(k_2+1)][l_1(l_1+1)+l_2(l_2+1)]}.\label{48}\nonumber\\
\end{eqnarray}
We compute
\begin{eqnarray}
(-1)^{\mu}Tr_L[L_{\mu}^{(1)},\hat{\mathcal{Y}}_{k_1n_1;k_2n_2}][A_{-\mu}^{(1)},\yyd_{l_1m_1;l_2m_2}]&=&\sum_{p_1s_1,p_2s_2}A_{-\mu}^{(1)}(p_1s_1p_2s_2)(-1)^{m_1+m_2}(-1)^{n_1+n_2}\sqrt{k_1(k_1+1)}\nonumber\\
&\times &\tilde{\Omega}_{p_1p_2,l_1l_2}^{k_1k_2}
C_{k_1n_11\mu}^{k_1n_1+\mu}C_{p_1s_1l_1-m_1}^{k_1-n_1-{\mu}}C_{p_2s_2l_2-m_2}^{k_2-n_2}\nonumber\\
(-1)^{\nu}Tr_L[L_{\nu}^{(1)},\hat{\mathcal{Y}}_{l_1m_1;l_2m_2}][A_{-\nu}^{(1)},\yyd_{k_1n_1;k_2n_2}]&=&\sum_{q_1t_1,q_2t_2}A_{-\nu}^{(1)}(q_1t_1q_2t_2)(-1)^{m_1+m_2}(-1)^{n_1+n_2}\sqrt{l_1(l_1+1)}\nonumber\\
&\times &\tilde{\Omega}_{q_1q_2,k_1k_2}^{l_1l_2}
C_{l_1m_11\nu}^{l_1m_1+\nu}C_{q_1t_1k_1-n_1}^{l_1-m_1-{\nu}}C_{q_2t_2k_2-n_2}^{l_2-m_2}
\end{eqnarray}
and
\begin{eqnarray}
(-1)^{\nu}Tr_L[A_{-\nu}^{(1)},\hat{\mathcal{Y}}_{l_1m_1;l_2m_2}][L_{\nu}^{(1)},\yyd_{k_1n_1;k_2n_2}]&=&\sum_{q_1t_1,q_2t_2}A_{-\nu}^{(1)}(q_1t_1q_2t_2)\sqrt{k_1(k_1+1)}\tilde{\Omega}_{q_1q_2,l_1l_2}^{k_1k_2}
C_{k_1-n_11\nu}^{k_1-n_1+\nu}C_{q_1t_1l_1m_1}^{k_1n_1-{\nu}}\nonumber\\
&\times &
C_{q_2t_2l_2m_2}^{k_2n_2}.
\end{eqnarray}
The sum over $n_2$ and $m_2$ can be done using the identity
\begin{eqnarray}
\sum_{n_2,m_2}C_{p_2s_2l_2-m_2}^{k_2-n_2}C_{q_2t_2k_2-n_2}^{l_2-m_2}=\frac{\sqrt{(2k_2+1)(2l_2+1)}}{2p_2+1}(-1)^{k_2+l_2+s_2}{\delta}_{q_2p_2}{\delta}_{t_2,-s_2}
\end{eqnarray}
whereas the sum over $n_1$ and $m_1$ can be done using

\begin{eqnarray}
\sum_{n_1,m_1}C_{k_1n_11\mu}^{k_1n_1+\mu}C_{p_1s_1l_1-m_1}^{k_1-n_1-\mu}C_{l_1m_11\nu}^{l_1m_1+\nu}C_{q_1t_1k_1-n_1}^{l_1-m_1-\nu}&=&\frac{\sqrt{(2k_1+1)(2l_1+1)}(-1)^{k_1+l_1+t_1+\mu}f_1(k_1l_1p_1s_1q_1t_1;\mu,\nu)}{\sqrt{k_1(k_1+1)l_1(l_1+1)(2p_1+1)(2q_1+1)}}\nonumber\\
\end{eqnarray}
where $f_1$ is the function which appears in the single fuzzy sphere case in equation $(C.9)$ of \cite{ref}. Explicitly it is given by

\begin{eqnarray}
f_1(k_1l_1p_1s_1q_1t_1;\mu,\nu)&=&\sqrt{k_1(k_1+1)l_1(l_1+1)(2p_1+1)(2q_1+1)(2k_1+1)(2l_1+1)}
\sum_{km}C_{p_1s_11\mu}^{km}C_{q_1t_11\nu}^{k-m}\nonumber\\
&\times &\left\{ \begin{array}{ccc}
                   l_1 & k_1 & p_1 \\
            1 & k  & k_1
                 \end{array} \right\}\left\{ \begin{array}{ccc}
                   k_1 & l_1 & q_1 \\
            1 & k  & l_1
                 \end{array} \right\}.
\end{eqnarray}
The first term of (\ref{48}) becomes
\begin{eqnarray}
2(L+1)^2\sum_{p_1s_1,p_2s_2}\sum_{q_1t_1}A_{-\mu}^{(1)}(p_1s_1p_2s_2)A_{-\nu}^{(1)}(q_1t_1p_2-s_2)(-1)^{s_1+s_2+\nu}&\times &\nonumber\\
\sum_{k_1,k_2}\sum_{l_1,l_2}\frac{(2k_1+1)(2k_2+1)}{k_1(k_1+1)+k_2(k_2+1)}\frac{(2l_1+1)(2l_2+1)}{l_1(l_1+1)+l_2(l_2+1)}
[1-(-1)^{R+p_1+p_2}][1-(-1)^{R+q_1+p_2}]&\times &\nonumber\\\left\{ \begin{array}{ccc}
                   p_1 & k_1 & l_1 \\
            \frac{L}{2} & \frac{L}{2}  & \frac{L}{2}
                 \end{array} \right\}\left\{ \begin{array}{ccc}
                   q_1 & k_1 & l_1 \\
            \frac{L}{2} & \frac{L}{2}  & \frac{L}{2}
                 \end{array} \right\}\left\{ \begin{array}{ccc}
                   p_2 & k_2 & l_2 \\
            \frac{L}{2} & \frac{L}{2}  & \frac{L}{2}
                 \end{array} \right\}^2f_1(k_1l_1p_1s_1q_1t_1;\mu,\nu).\label{54}
\end{eqnarray}
Next we compute the second term of (\ref{48}). The sum over $n_1$ and $m_1$ will now be done using the identity

\begin{eqnarray}
\sum_{n_1,m_1}(-1)^{n_1+m_1}C_{k_1n_11\mu}^{k_1n_1+\mu}C_{k_1-n_11\nu}^{k_1-n_1+\nu}C_{p_1s_1l_1-m_1}^{k_1-n_1-\mu}C_{q_1t_1l_1m_1}^{k_1n_1-\nu}&=&\frac{(2k_1+1)(-1)^{s_1+\nu}f_2(k_1l_1p_1s_1q_1t_1;\mu,\nu)}{k_1(k_1+1)\sqrt{(2p_1+1)(2q_1+1)}}\nonumber\\
\end{eqnarray}
where $f_2$ is the other function which appears in the single fuzzy sphere case in equation $(C.14)$ of \cite{ref}. Explicitly it is given by

\begin{eqnarray}
f_2(k_1l_1p_1s_1q_1t_1;\mu,\nu)&=&k_1(k_1+1)(2k_1+1)\sqrt{(2p_1+1)(2q_1+1)}
\sum_{km}(-1)^{k+k_1+l_1}C_{p_1s_11\mu}^{km}C_{q_1t_11\nu}^{k-m}\nonumber\\
&\times &\left\{ \begin{array}{ccc}
                   l_1 & k_1 & p_1 \\
            1 & k  & k_1
                 \end{array} \right\}\left\{ \begin{array}{ccc}
                   l_1 & k_1 & q_1 \\
            1 & k  & k_1
                 \end{array} \right\}.
\end{eqnarray}
We obatin the same result (\ref{54}) with the replacement $f_1{\longrightarrow}(-1)^{k_2+l_2+p_2}f_2$ and hence the full $3-$vertex correction will be given by

\begin{eqnarray}
2(L+1)^2\sum_{p_1s_1,p_2s_2}\sum_{q_1t_1}A_{-\mu}^{(1)}(p_1s_1p_2s_2)A_{-\nu}^{(1)}(q_1t_1p_2-s_2)(-1)^{s_1+s_2+\nu}&\times &\nonumber\\
\sum_{k_1,k_2}\sum_{l_1,l_2}\frac{(2k_1+1)(2k_2+1)}{k_1(k_1+1)+k_2(k_2+1)}\frac{(2l_1+1)(2l_2+1)}{l_1(l_1+1)+l_2(l_2+1)}
[1-(-1)^{R+p_1+p_2}][1-(-1)^{R+q_1+p_2}]&\times &\nonumber\\\left\{ \begin{array}{ccc}
                   p_1 & k_1 & l_1 \\
            \frac{L}{2} & \frac{L}{2}  & \frac{L}{2}
                 \end{array} \right\}\left\{ \begin{array}{ccc}
                   q_1 & k_1 & l_1 \\
            \frac{L}{2} & \frac{L}{2}  & \frac{L}{2}
                 \end{array} \right\}\left\{ \begin{array}{ccc}
                   p_2 & k_2 & l_2 \\
            \frac{L}{2} & \frac{L}{2}  & \frac{L}{2}
                 \end{array} \right\}^2&\times&\nonumber\\
\bigg(f_1(k_1l_1p_1s_1q_1t_1;\mu,\nu)+(-1)^{k_2+l_2+p_2}f_2(k_1l_1p_1s_1q_1t_1;\mu,\nu)\bigg).
\end{eqnarray}
Let us remark that we must have the conservation laws $R+p_1+p_2={\rm odd}$ and $R+q_1+p_2={\rm odd}$ and hence we must always have  $p_1+q_1={\rm even}$. In $f_1$ and $f_2$ the angular momentum $k$ can only take the values $k=p_1$,$k=p_1+1$ and $k=p_1-1$ or equivalently $k=q_1$,$k=q_1+1$ and $k=q_1-1$. Thus there is only one term in $f_1+(-1)^{k_2+l_2+p_2}f_2$ in which $q_1=p_1$ given by
\begin{eqnarray}
\bigg(\sqrt{k_1(k_1+1)l_1(l_1+1)(2p_1+1)(2q_1+1)(2k_1+1)(2l_1+1)}\left\{ \begin{array}{ccc}
                   l_1 & k_1 & p_1 \\
            1 & p_1  & k_1
                 \end{array} \right\}\left\{ \begin{array}{ccc}
                   k_1 & l_1 & p_1 \\
            1 & p_1  & l_1
                 \end{array} \right\}&-&\nonumber\\k_1(k_1+1)(2k_1+1)\sqrt{(2p_1+1)(2q_1+1)}
\left\{ \begin{array}{ccc}
                   l_1 & k_1 & p_1 \\
            1 & p_1  & k_1
                 \end{array} \right\}\left\{ \begin{array}{ccc}
                   l_1 & k_1 & p_1 \\
            1 & p_1  & k_1
                 \end{array} \right\}\bigg)C_{p_1s_11\mu}^{p_1m}C_{q_1t_11\nu}^{p_1-m}{\delta}_{p_1q_1}\nonumber\\
\end{eqnarray}
This term leads to the contribution ( by using the tables on page $311$ of \cite{VKM} )
\begin{eqnarray}
\sum_{p_1s_1,p_2s_2}A_{-\mu}^{(1)}(p_1s_1p_2s_2)A_{-\nu}^{(1)}(p_1-m-\nu p_2-s_2)(-1)^{s_1+s_2+\nu}C_{p_1s_11\mu}^{p_1m}C_{p_1-m-\nu 1\nu}^{p_1-m}p_1(p_1+1){\cal O}_3(p_1,p_2)&=&\nonumber\\
-Tr_L{\cal L}_a^{(1)}A_a^{(1)}{\cal O}_3({\Delta}_1,{\Delta}_2){\cal L}_b^{(1)}A_b^{(1)}.\nonumber\\
\end{eqnarray}
where
\begin{eqnarray}
{\cal O}_3(p_1,p_2)&=&-4(L+1)^2
\sum_{k_1,k_2}\sum_{l_1,l_2}\frac{(2k_1+1)(2k_2+1)}{k_1(k_1+1)+k_2(k_2+1)}\frac{(2l_1+1)(2l_2+1)}{l_1(l_1+1)+l_2(l_2+1)}\nonumber\\
&\times &[1-(-1)^{R+p_1+p_2}]\left\{ \begin{array}{ccc}
                   p_1 & k_1 & l_1 \\
            \frac{L}{2} & \frac{L}{2}  & \frac{L}{2}
                 \end{array} \right\}^2\left\{ \begin{array}{ccc}
                   p_2 & k_2 & l_2 \\
            \frac{L}{2} & \frac{L}{2}  & \frac{L}{2}
                 \end{array} \right\}^2
\frac{k_1(k_1+1)\big(l_1(l_1+1)-k_1(k_1+1)\big)}{p_1^2(p_1+1)^2}.\label{61}\nonumber\\
\end{eqnarray}
The remaining terms has the structure
\begin{eqnarray}
C_{p_1s_11\mu}^{p_1+1 m}C_{q_1t_11\nu}^{p_1+1-m}{\eta}_{+}(k_1l_1;p_1q_1)+C_{p_1s_11\mu}^{p_1-1 m}C_{q_1t_11\nu}^{p_1-1-m}{\eta}_{-}(k_1l_1;p_1q_1)
\end{eqnarray}
where
\begin{eqnarray}
{\eta}_{\pm}(k_1l_1;p_1q_1)=\sqrt{(2p_1+1)(2q_1+1)k_1(k_1+1)(2k_1+1)}\left\{ \begin{array}{ccc}
                   l_1 & k_1 & p_1 \\
            1 & p_1\pm 1  & k_1
                 \end{array} \right\}&\times &\nonumber\\
\bigg(\sqrt{l_1(l_1+1)(2l_1+1)}\left\{ \begin{array}{ccc}
                   k_1 & l_1 & q_1 \\
            1 & p_1\pm 1  & l_1
                 \end{array} \right\}
+\sqrt{k_1(k_1+1)(2k_1+1)}\left\{ \begin{array}{ccc}
                   l_1 & k_1 & q_1 \\
            1 & p_1\pm 1  & k_1
                 \end{array} \right\}\bigg).\nonumber\\
\end{eqnarray}
We have the final contributions
 \begin{eqnarray}
\sum_{p_1s_1,p_2s_2}\sum_{q_1t_1}A_{-\mu}^{(1)}(p_1s_1p_2s_2)A_{-\nu}^{(1)}(q_1t_1 p_2-s_2)(-1)^{s_1+s_2+\nu}C_{p_1s_11\mu}^{p_1{\pm}1m}C_{q_1t_1 1\nu}^{p_1{\pm}1-m}{\Sigma}_{\pm}(p_1,p_2)\label{74}
\end{eqnarray}
 where
\begin{eqnarray}
{\Sigma}_{\pm}&=&2(L+1)^2
\sum_{k_1,k_2}\sum_{l_1,l_2}\frac{(2k_1+1)(2k_2+1)}{k_1(k_1+1)+k_2(k_2+1)}\frac{(2l_1+1)(2l_2+1)}{l_1(l_1+1)+l_2(l_2+1)}\nonumber\\
&\times &[1-(-1)^{R+p_1+p_2}][1-(-1)^{R+q_1+p_2}]\left\{ \begin{array}{ccc}
                   p_1 & k_1 & l_1 \\
            \frac{L}{2} & \frac{L}{2}  & \frac{L}{2}
                 \end{array} \right\}\left\{ \begin{array}{ccc}
                   q_1 & k_1 & l_1 \\
            \frac{L}{2} & \frac{L}{2}  & \frac{L}{2}
                 \end{array} \right\}\left\{ \begin{array}{ccc}
                   p_2 & k_2 & l_2 \\
            \frac{L}{2} & \frac{L}{2}  & \frac{L}{2}
                 \end{array} \right\}^2{\eta}_{\pm}\nonumber\\\label{75}
\end{eqnarray}
It is not difficult to show that the contributions (\ref{74}) will involve anticommutators between $A_a^{(1)}$ and $L_a^{(1)}$ instead of commutators. Hence it is of the same type as the scalar action 
\begin{eqnarray}
Tr_L[L_a,A_a]_{+}^2
\end{eqnarray}
Indeed we have shown in \cite{ref} that (\ref{74}) ( or more precisely the analogue of (\ref{74}) for a single fuzzy ${\bf S}^2$ ) is the sum of four terms each of the form
\begin{eqnarray}
-Tr_L[V_i(A_a^{(1)}),L_a^{(1)}]{\Delta}_{ij}({\Delta}_1,{\Delta}_2)[V_j(A_b^{(1)}),L_b^{(1)}]\label{77}
\end{eqnarray}
Following the same method used in reference \cite{ref} we can give explicit expressions for the operators $V_i$ and ${\Delta}_{ij}$ by comparing (\ref{74}) and (\ref{75}) from one hand and (\ref{77}) from the other hand.

\end{document}